\documentclass[journal]{IEEEtran} 

\usepackage{amsmath,amsfonts,amssymb,amsthm} 
\usepackage{booktabs,tabularx}
\usepackage{array}
\usepackage{textcomp}
\usepackage{url}
\usepackage{verbatim}
\usepackage{graphicx}
\usepackage{cite}
\usepackage{mathrsfs}  
\usepackage{bm}
\usepackage{multirow}
\usepackage{enumitem}
\usepackage{bbm}
\usepackage{xcolor}

\usepackage[caption=false,font=footnotesize]{subfig}
\usepackage[colorlinks,
linkcolor=blue,       
urlcolor=blue,
citecolor=blue,       
]{hyperref}
\def\BibTeX{{\rm B\kern-.05em{\sc i\kern-.025em b}\kern-.08em
T\kern-.1667em\lower.7ex\hbox{E}\kern-.125emX}}
\usepackage{balance}
\usepackage{ragged2e}
\usepackage{tikz}
\newtheorem{definition}{Definition}

\newtheorem{proposition}{Proposition}
\newtheorem{example}{Example}

\usepackage{booktabs}
\usepackage{longtable}
\usepackage{tabularx}
\usepackage{amsmath, amssymb}
\usepackage{pifont}

\begin{document}
\title{
Spatiotemporal Feature Alignment and Weighted Fusion in Collaborative Perception Enabled by Network Synchronization and Age of Information
}
\author{Qiaomei~Han,~\IEEEmembership{Member,~IEEE,}
Xianbin~Wang,~\IEEEmembership{Fellow,~IEEE,}
and~Minghui~Liwang,~\IEEEmembership{Senior Member,~IEEE}
\thanks{

Qiaomei Han and Xianbin Wang are with the Department of Electrical and Computer Engineering, Western University, London, ON, N6A 5B9, Canada (e-mails: {qhan42, xianbin.wang}@uwo.ca).

Minghui Liwang is with the Department of Control Science and Engineering,  Tongji University, Shanghai, China  (email: minghuiliwang@tongji.edu.cn).

\textit{Corresponding author: Dr. Xianbin Wang.}}

}
\markboth{IEEE Transactions on Cognitive Communications and Networking}
{Han \MakeLowercase{\textit{et al.}}: Spatiotemporal Feature Alignment and Weighted Fusion in Collaborative Perception Enabled by Network Synchronization and Age of Information}
\maketitle
\begin{abstract}  
Collaborative perception in Internet of Vehicles (IoV) aggregates multi-vehicle observations for broader scene coverage and improved decision-making. However, fusion quality degrades under spatiotemporal heterogeneity from unsynchronized clocks, communication delays, and motion variations across vehicles. Prior work mitigates these through spatial transformations or fixed time-offset corrections, overlooking time-varying clock drifts and delays that cause persistent feature misalignment. To address these challenges, we propose a spatiotemporal feature alignment and weighted fusion framework. Specifically, network synchronization is introduced to estimate inter-vehicle clock states and establish a common temporal reference, onto which local feature timestamps can be mapped. Based on this, we define delivery-time Age of Information (AoI) to measure the expected age of a shared feature when it becomes available for fusion, by accounting for its generation time and the Vehicle-to-Everything (V2X)  communication delay. The proposed spatiotemporal feature alignment then compensates asynchronous neighbor features toward the fusion time, rather than directly aggregating delayed features. Since different spatial regions contribute unequally to perception, we further perform Region-of-Interest (RoI)-level weighted fusion, where the fusion weights are determined by delivery-time AoI, synchronization reliability, and content complementarity. As a result, timely, reliable, and complementary regions are emphasized, while stale, uncertain, or redundant regions are down-weighted. Simulation results further demonstrate consistent accuracy improvements over representative baselines under clock drift, varying communication conditions, temporal misalignment levels, and vehicle numbers.
\end{abstract}


\section{Introduction}

The proliferation of smart devices and distributed sensing platforms has driven rapid advances in collaborative computing for enhanced decision making and computation efficiency~\cite{9795129}. Representative techniques include federated learning~\cite{mcmahan2017communication}, split learning~\cite{han2021accelerating}, and transfer learning~\cite{9723467}, where distributed nodes process local data and exchange feature information across interconnected systems. Such collaborative computing paradigms have been widely applied in smart manufacturing, smart cities, and smart healthcare~\cite{9346022,10637735}.

Building on this paradigm, collaborative perception aggregates observations from multiple vehicles to obtain a broader and more accurate view of the driving environment, thereby improving decision making in Internet of Vehicles (IoV) systems~\cite{9606821,10752404,9057672}. Recent work increasingly focuses on feature-level collaboration, where vehicles exchange compact intermediate representations instead of raw sensor data or final detected objects. Compared with object-level collaboration, which discards contextual information, and raw-level collaboration, which incurs high bandwidth and latency costs, feature-level collaboration provides a practical balance between perceptual information and communication efficiency. Therefore, we target feature-level collaboration  in this work.

Despite its potential, feature-level collaborative perception often suffers from  spatiotemporal feature misalignment across vehicles. Spatial misalignment arises because features are generated in different local coordinate frames, whereas temporal misalignment is caused by unsynchronized clocks, sensing time differences, and communication delays~\cite{9943796,10649634}.
As a result, features from different vehicles correspond to different scene states and viewpoints when they are fused, directly aggregating them is ineffective and ultimately degrades the perception accuracy~\cite{9228884,9963987}.

To mitigate feature misalignment, current approaches  mainly address spatial domain variation. They transform features from different vehicles into a common coordinate system using rotation, translation, or learned geometric mapping techniques~\cite{8885377,xu2022v2x}.  Although these methods substantially reduce motion- and viewpoint-induced discrepancies, they rectify geometry only at a single timestamp, leaving temporal inconsistency unaddressed. Therefore, residual time offsets persist and manifest as apparent spatial domain drift over time, undermining alignment accuracy in dynamic scenes.

Beyond purely spatial alignment, recent studies have started to address temporal compensation. SyncNet~\cite{lei2022latency} predicts the current feature map from past inputs under a latency-aware collaborative perception framework, while CoBEVFlow~\cite{wei2023asynchrony} estimates a Bird's-Eye-View (BEV) flow field to warp asynchronous features from their source timestamps toward the receiver time. However, these methods usually assume that the timestamps used for compensation are already comparable across vehicles. In practical inter-vehicle collaboration, independent vehicle clocks often suffer from offset and skew~\cite{9943796,10649634}, while communication latency varies over time. As a result, local timestamps cannot be directly used as compensation references and should be mapped onto a common temporal reference. This mapping determines the synchronized interval from feature generation to fusion, then used for temporal compensation. Nevertheless, this mapping is still an estimate, and its residual uncertainty can reduce the temporal reliability, thereby degrading alignment and perception performance.

Moreover, to improve communication and fusion efficiency, Region-of-Interest (RoI)-level weighted fusion has been investigated to prioritize informative regions instead of aggregating full feature maps~\cite{liu2020who2com,hu2022where2comm}.
Recent studies have introduced information freshness into RoI prioritization, commonly using Age of Information (AoI) to characterize the age of the freshest update available at the receiver, measured with respect to the generation time of that update~\cite{yates2021aoi_survey}. Studies on 6G and intelligent vehicular networks further indicate that freshness should be placed in a service-oriented timing framework that relates the generation, delivery, and application of information to the service objective~\cite{popovski2022perspective_time,guo2023aoi_vehicular}. However, in RoI-level collaborative perception,  RoI prioritization is often performed before candidate RoI features are transmitted. Since these features continue to age during inter-vehicle communication, their freshness should be evaluated according to the expected age  at the fusion time.

Furthermore, freshness alone is insufficient for RoI-level weighted fusion. A low AoI only indicates that the shared RoI is timely, but does not guarantee that it is temporally reliable or useful for fusion. As discussed above, residual synchronization uncertainty can affect temporal reference construction and spatiotemporal feature alignment~\cite{9590496}. In addition, the contribution of a shared RoI also depends on whether it provides complementary evidence for collaborative perception. If a shared RoI largely overlaps with the ego observation or contains limited task-relevant information, its benefit can be reduced~\cite{giordani2019voi,lyu2025accuracy,wolff2025uncertainty}. Therefore, we formulate the fusion weights by jointly considering delivery-time freshness, synchronization reliability, and content complementarity.

Motivated by these, we propose a spatiotemporal feature alignment and weighted fusion framework for asynchronous inter-vehicle collaborative perception. The main contributions are summarized as follows:
\begin{itemize}
\item To establish an accurate temporal reference, we design network synchronization to estimate and update inter-vehicle clock states. This enables local feature timestamps to be mapped onto a  common temporal reference and provides residual synchronization uncertainty for reliability assessment. On this temporal reference, delivery-time AoI is defined to evaluate the freshness of candidate RoIs at their expected available time for fusion.
\item Based on the temporal reference, we propose a spatiotemporal
feature alignment mechanism. It first uses geometric projection to transform
neighbor features into a common fusion frame, so that temporal compensation can
be performed under a consistent spatial representation. The projected features
are then adjusted according to the feature age computed from synchronized
timestamps, mitigating delay-induced feature mismatch before fusion. This design improves the space-time consistency of asynchronous features in
collaborative perception.
\item To improve communication and fusion efficiency, we formulate an RoI-level fusion utility that jointly considers delivery-time AoI, synchronization reliability, and content complementarity. This utility guides RoI selection and fusion weighting, prioritizing fresh, reliable, and complementary regions while reducing the contribution of stale, uncertain, or
redundant ones.
\item The proposed framework is trained end-to-end and optimized with detection, alignment, and objectness objectives. Experiments demonstrate consistent accuracy improvements over representative baselines under clock drift, varying communication conditions, temporal misalignment levels, voxel sizes,  backbone architectures and vehicle numbers. Ablation studies further verify the effectiveness of the major components. 
\end{itemize}

The remainder of this article is organized as follows:  Section II reviews the related work on feature alignment and  weighted fusion in collaborative perception. Section III describes our proposed spatiotemporal feature alignment and weighted fusion framework by network synchronization  and AoI. To evaluate its effectiveness, Section IV demonstrates simulation results. Finally, Section V  concludes the paper and proposes  future directions.

\section{Related Work} 
In this section, we review the existing work related to feature alignment and fusion, overcoming the challenges arised from spatiotemporal heterogeneity and inefficiency.

\subsection{Spatiotemporal Feature Alignment}
Due to spatiotemporal heterogeneity across vehicles caused by unsynchronized clocks, network delays, or motion discrepancies, the quality of collaborative perception is degraded. Early approaches mainly mitigate spatial feature misalignment by projecting features or detections into a common coordinate system  via rigid rotations, translations or by warping BEV grids to compensate for viewpoint changes. NEAT~\cite{yang2024align} predicts feature-level corrections prior to fusion, improving robustness to pose noise.  Vehicle-to-Everything (V2X)-ViT~\cite{xu2022v2x}   applies an attention mechanism to aggregate cross-agent information, effectively learning where and how to fuse. These designs reduce geometric mismatch but presume synchronous inputs. When agents are time-shifted by latency or clock drift, they cannot eliminate the temporal misalignment.

Another line of work targets temporal feature  alignment. For example, SyncNet~\cite{lei2022latency} compensates latency by estimating asynchronous features at a common timestamp through feature-attention-based estimation and time modulation. CoBEVFlow~\cite{wei2023asynchrony} estimates BEV flow to relocate asynchronous sender features from their source timestamps toward the receiver time, improving collaborative perception under irregular temporal offsets. However, these methods often assume that timestamps used for temporal compensation are synchronized and directly comparable across vehicles. This assumption is difficult to satisfy in practical IoV systems, where independent vehicle clocks exhibit offset and skew. Directly using local timestamps can lead to imperfect temporal compensation, especially for dynamic objects that are sensitive to temporal misalignment~\cite{10891245}.

Therefore, network synchronization is essential to track inter-vehicle clock states and map asynchronous feature timestamps onto a common temporal reference before feature alignment. Since this timestamp mapping is obtained from estimated clock states, residual timing uncertainty remains and should also be considered when evaluating the temporal reliability of shared features.

\subsection{Weighted Feature Fusion}
Communication-efficient collaborative perception often prioritizes informative RoIs rather than transmitting full feature maps. Where2Comm~\cite{hu2022where2comm} learns a spatial confidence map to identify critical regions for communication. When2com~\cite{liu2020when2com} learns when communication should be activated and how communication groups should be formed. Who2com~\cite{liu2020who2com} determines which agents should be queried when the ego observation becomes unreliable. These methods mainly rely on perceptual confidence or communication decisions to select RoIs.  Nevertheless, in asynchronous collaborative perception, the utility of a candidate RoI also depends on freshness, since features generated at different times can correspond to different scene states when used for perception.

Information freshness is commonly characterized by AoI, which measures the elapsed time since the generation of the most recently received update~\cite{yates2021aoi_survey}. AoI and its variants, such as time-average AoI, peak AoI, and AoI violation probability, have been widely used to describe timeliness in networked and vehicular systems~\cite{yates2021aoi_survey, guo2023aoi_vehicular}. In collaborative perception, Fresh2Comm~\cite{wu2025fresh2comm} incorporates AoI into feature prioritization, showing the need to distinguish stale perception features from recently generated ones.

For weighted fusion, freshness should correspond to the age of a candidate RoI when it becomes available for fusion. Conventional AoI characterizes the age of the freshest update available at the receiver, measured with respect to the generation time of that update~\cite{yates2021aoi_survey}. However, in collaborative perception, RoI prioritization is often performed before the corresponding features reach the receiving vehicle. As they continue to age during inter-vehicle communication, freshness assessed before transmission does not capture their expected staleness when used for fusion.  Recent 6G studies also indicate that freshness should be placed in a service-oriented timing framework that relates the generation, delivery, and application of information to the service objective~\cite{popovski2022perspective_time}. Therefore, we define delivery-time AoI to quantify the expected age of a candidate RoI at the fusion time.

Recent AoI-aware studies in vehicular networks further show that information freshness is closely related to communication dynamics and perception-oriented information selection. In V2X networks, resource contention, interference, and channel degradation can increase the access and transmission delay of shared information~\cite{rolich2025aoi_reuse}. Accordingly, Mlika et al.~\cite{mlika2022ddpg_aoi_v2x} reduced AoI through joint resource and power allocation,  improving information timeliness under wireless resource constraints. Zhu et al.~\cite{zhu2026tamp} considered AoI together with communication volume in multi-region collaborative perception, reducing stale regional updates while limiting communication overhead. Fang et al.~\cite{fang2024racp} introduced Age of Perceived Targets (AoPT) and integrated target-level freshness into task-oriented compression, reducing redundant information exchange while preserving perception-relevant targets. These studies motivate evaluating RoI freshness according to not only when information is generated, but also when it becomes available and how it supports the perception task.

Beyond freshness, RoI-level weighted fusion also needs to evaluate whether a candidate RoI provides reliable and non-redundant information to the receiving vehicle. Giordani et al.~\cite{giordani2019voi} studied Value of Information (VoI) in vehicular networks and evaluated shared information using spatial, temporal, and quality-related attributes. Lyu et al.~\cite{lyu2025accuracy} prioritized perceived objects from the perspective of accuracy and relevance, while Wolff et al.~\cite{wolff2025uncertainty} incorporated uncertainty and redundancy into information selection for Collective Perception Messages (CPMs). In addition, Wang et al.~\cite{wang2024aoci_v2x} introduced content-aware AoI to incorporate information-content changes into freshness evaluation. These studies suggest that useful shared information should be evaluated beyond freshness alone, with temporal reliability and content complementarity also considered. However, they mainly assign value at the object or message level, rather than providing feature-level fusion weights for perception. Accordingly, we formulate a  weighted fusion utility that combines delivery-time freshness, synchronization reliability, and content complementarity to guide feature aggregation.

\section{A Spatiotemporal Feature Alignment and Weighted Fusion Framework}
\begin{figure}[!t]
\centering \includegraphics[width=3.3in]{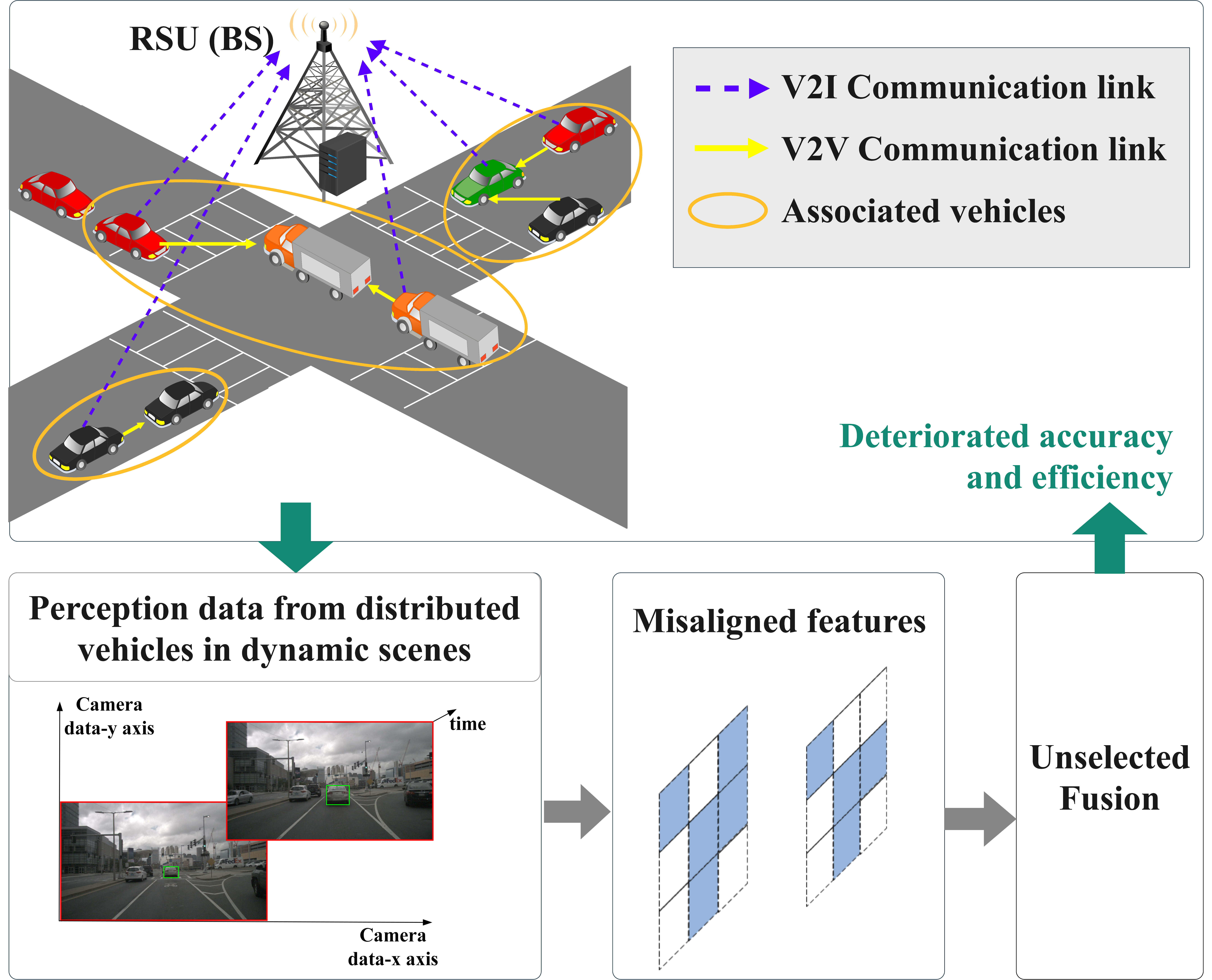}
\caption{Collaborative perception impacted by spatiotemporal feature misalignment and inefficient feature fusion.}
\label{fig_1}
\end{figure}

Collaborative perception can improve scene understanding by aggregating observations from multiple vehicles. However, its accuracy and efficiency are often hindered by spatiotemporal feature misalignment and inefficient feature fusion, as shown in Fig.~\ref{fig_1}. To address these challenges, we introduce a spatiotemporal feature alignment and weighted fusion framework, as detailed below.

\subsection{System Model}
We consider an IoV system comprising a set of vehicles $\mathcal{M}=\{1,\ldots,M\}$. Each vehicle is equipped with an onboard unit that supports local computation and wireless communication. A roadside unit (RSU), integrating an edge server and a base station (BS), operates within the coverage area and supports collaborative perception.

Let $\psi_r$ and $\psi_v$ denote the RSU coverage radius and the lateral offset between the road axis and the RSU center, respectively. Then the half-length of the RSU coverage region projected onto the road axis is $\sqrt{\psi_r^2-\psi_v^2}$, assuming $|\psi_v|<\psi_r$. For vehicle $m$, its remaining distance before leaving the RSU coverage region is  $d_m = \sqrt{\psi_r^2-\psi_v^2} -{v_m}/{|v_m|} \cdot \psi_m$,   where $\psi_m$ is the position of vehicle $m$ along the road axis, $v_m$ is its velocity, and $v_m/|v_m|$ indicates its driving direction. The neighbor set of vehicle $m$ is thus defined as \begin{equation} 
{L}_m = \left\{ l \in \mathcal{M}\setminus\{m\} \,\middle|\, \left\| \boldsymbol{\psi}_l-\boldsymbol{\psi}_m \right\| \le \iota \right\}, 
\end{equation} 
where $\iota$ is the communication range.

\subsubsection{Feature Extraction}
For the ego vehicle $m$, its LiDAR sensor collects point-cloud data denoted by 
\begin{equation} \boldsymbol{X}_m = \left\{ \boldsymbol{p}_{m,i} = (x_{m,i},y_{m,i},z_{m,i},\rho_{m,i}) \mid i=1,\ldots,N_m \right\}, 
\end{equation} 
where $(x_{m,i},y_{m,i},z_{m,i})$ is the 3D coordinate of the $i$-th point, $\rho_{m,i}$ is its intensity, and $N_m$ is the number of collected points. The point cloud is voxelized into pillars and encoded by a PointPillars-based encoder~\cite{lang2019pointpillars}, yielding the BEV feature map 
\begin{equation} 
\boldsymbol{F}_m(t) = \Phi_{\mathrm{enc}} \left( \boldsymbol{X}_m(t) \right), 
\end{equation} 
where $\Phi_{\mathrm{enc}}(\cdot)$ denotes the encoder and $\boldsymbol{F}_m(t)\in\mathbb{R}^{X'\times Y'\times C}$ is the BEV feature representation.

All vehicles use the same encoder. Neighboring vehicles generate BEV features from their LiDAR observations and attach the corresponding feature-generation timestamps and synchronization metadata. The ego feature is kept as the fusion anchor, while the received neighbor features are subsequently mapped, compensated, and fused in the ego fusion frame.

Building on this paradigm, we introduce a spatiotemporal feature alignment and weighted fusion framework, as illustrated in Fig.~\ref{fig_2}. First, network synchronization establishes a common temporal reference among vehicles, supporting synchronized timestamp mapping,  AoI and reliability calculation. Second, spatiotemporal feature alignment maps neighboring features to the ego fusion frame through geometric projection and AoI-conditioned feature compensation. Third, RoI-level weighted fusion aggregates the compensated features by assigning larger utilities to regions that are fresh, reliable, and complementary to the ego feature.

\begin{figure*}[!t]
\centering \includegraphics[width=7in]{Framework.eps}
\caption{Workflow of the proposed collaborative perception framework, including network synchronization, spatiotemporal feature alignment, and weighted feature fusion.}
\label{fig_2}
\end{figure*}

\subsubsection{Message Architecture}
Inspired by the ETSI ITS communication architecture for collective perception~\cite{etsi_en302637,etsi_ts103324}, we consider three types of messages in the proposed framework:
\begin{itemize}
\item \textbf{Synchronization messages~\cite{9943796}:} These messages support bidirectional timestamp exchange between vehicles. They are used to estimate and update the relative clock state, providing a common temporal reference and synchronization uncertainty for freshness and reliability evaluation.
\item \textbf{Cooperative Awareness Messages (CAMs)~\cite{etsi_en302637}:} CAMs periodically report vehicle states, such as position, heading, and velocity. These state attributes provide the mobility and pose information needed for geometric projection and spatiotemporal alignment.
\item \textbf{Collective Perception Messages (CPMs)~\cite{etsi_ts103324}:}
CPMs provide a message structure for sharing perception information in collective perception. In this work, we use a feature-level CPM representation to describe selected RoI feature payloads and the associated timing and content metadata required by the fusion module. The detailed structure is specified in Eq.~\eqref{eq:fcpm}.
\end{itemize}

\subsection{Network Synchronization}
\label{sec:clock-sync}
Without accurate synchronization, vehicles may assign different timestamps to the same physical event, leading to inconsistent feature ages and mismatched feature fusion. Therefore, for ego vehicle $m$, we maintain a relative clock state with respect to each neighboring vehicle $l\in{L}_m$. This clock state provides a common temporal reference for subsequent alignment and fusion operations.

\subsubsection{Two-Way Timestamp Exchange}
The ego vehicle $m$ estimates its clock relation with each neighboring vehicle $l\in{L}_m$ through a two-way timestamp exchange procedure based on IEEE~1588 Precision Time Protocol (PTP)~\cite{9943796}. At the $k$-th exchange round, the timestamp exchange between vehicles $m$ and $l$ produces six timestamps: 
\begin{itemize} 
\item $m$ sends a \texttt{Sync} message to $l$ at time $t_{1,k}$ according to $m$'s clock, and $l$ receives it at $t_{2,k}$ according to $l$'s clock; 
\item $l$ sends a \texttt{Delay\_Req} message at time $t_{3,k}$ based on $l$'s clock, and $m$ receives it at $t_{4,k}$ according to $m$'s clock; 
\item $l$ sends another timestamped message at   $t_{5,k}$ according to $l$'s clock, and $m$ receives it at $t_{6,k}$  based on  $m$'s clock. 
\end{itemize}

Vehicle $l$ then reports its recorded timestamps $\{t_{2,k},t_{3,k},t_{5,k}\}$ to vehicle $m$. Under the symmetric one-way delay assumption, coarse observations of the clock offset and relative clock skew  can be obtained as 
\begin{align} 
\tilde{\theta}_k &= \frac{(t_{2,k}-t_{1,k})-(t_{4,k}-t_{3,k})}{2},
\label{eq:coarse-offset}  \\  \tilde{\varpi}_k &= \frac{t_{5,k}-t_{3,k}}{t_{6,k}-t_{4,k}} -1. 
\label{eq:coarse-skew} 
\end{align}

However, in dynamic V2X environments, forward and reverse delays are generally asymmetric and time-varying,  which perturbs timestamp observations and introduces bias into the clock-state estimates. Therefore, a recursive estimator is needed to track the clock relation over consistent timestamp exchanges and to quantify the synchronization reliability. 

\subsubsection{Clock State Model and Recursive Estimation}
\label{sec:clock-state-estimation}

Although Eqs.~\eqref{eq:coarse-offset}--\eqref{eq:coarse-skew} provide coarse observations of the clock relation, building a common temporal reference requires tracking the clock state across exchange rounds. We define the clock state as
$\boldsymbol{x}_k=[\theta_k\ \varpi_k]^\top$, where $\theta_k$ is the clock offset and $\varpi_k$ is the relative clock skew. The offset represents the time difference between the local clock and the reference clock at a given change round, whereas the skew characterizes the relative clock-rate mismatch, which causes the offset to drift between synchronization updates. Therefore, the two-state formulation is used to maintain timestamp consistency under possibly irregular synchronization updates.

We write the coarse observation vector as $\boldsymbol{z}_k=[\tilde{\theta}_k\ \tilde{\varpi}_k]^\top$, where $\tilde{\theta}_k$ and $\tilde{\varpi}_k$ are obtained from Eqs.~\eqref{eq:coarse-offset}--\eqref{eq:coarse-skew}. Due to timestamp perturbations and forward--reverse delay asymmetry, these observations are biased and noisy versions of the clock state.  The observation model can be formulated as~\cite{10.1109/TNET.2011.2158656} 
\begin{align}
&	\boldsymbol{z}_k=
\boldsymbol{A}\boldsymbol{x}_k
+	\boldsymbol{H} b_k
+	\boldsymbol{\eta}_k, \\ \notag
&	\text{s.t.} \
\boldsymbol{A}
=	\begin{bmatrix}
1 & 0\\
0 & 1
\end{bmatrix},
\ 	\boldsymbol{H}=[1\ 0]^\top,
\label{eq:A-Hb-sync}
\end{align} 
where $b_k$ denotes the asymmetry-induced bias in the current timestamp
exchange. With $\boldsymbol{H}=[1\ 0]^\top$, this bias affects the coarse offset observation in this model.  $\boldsymbol{\eta}_k$ denotes the timestamp-observation noise caused by delay fluctuation, jitter, and other random disturbances.  Its covariance
$\boldsymbol{I}_k=\mathbb{E}[\boldsymbol{\eta}_k\boldsymbol{\eta}_k^\top]
\in\mathbb{R}^{2\times 2}$  captures the noise variances of the coarse offset and skew observations, alongside their cross-covariance.

Over the interval $\Delta t_k=t_k-t_{k-1}$, the clock state evolves 
\begin{equation} 
\boldsymbol{x}_k = \boldsymbol{\Phi}_k\boldsymbol{x}_{k-1} + \boldsymbol{w}_k, \label{eq:clock-state-transition} 
\end{equation} 
with the state-transition matrix  $
\boldsymbol{\Phi}_k
=
\begin{bmatrix}
1 & \Delta t_k\\
0 & 1
\end{bmatrix}$.
It describes the offset evolution caused by relative clock skew over the elapsed interval. The process noise $\boldsymbol{w}_k$ captures temporal variations in the clock offset and relative clock skew, and is modeled as zero-mean Gaussian noise with covariance 
\begin{equation} 
\boldsymbol{Q}_{d,k} = 
\begin{bmatrix} q_\theta\Delta t_k+\frac{1}{3}q_\varpi\Delta t_k^3 & \frac{1}{2}q_\varpi\Delta t_k^2\\ \frac{1}{2}q_\varpi\Delta t_k^2 & q_\varpi\Delta t_k 
\end{bmatrix}, 
\label{eq:Qd-k-revised} 
\end{equation} 
where $q_\theta$ and $q_\varpi$ control the intensities of the offset noise and relative-skew noise, respectively. 

The above model separates the clock state $(\theta_k,\varpi_k)$ from the asymmetry-induced bias $b_k$. A standard Kalman filter can recursively track clock state~\cite{10684400}, but it does not explicitly compensate for the bias introduced by asymmetric delays. To reduce the impact of asymmetry-induced bias on the clock-state update, we adopt the Robust Three-Step Recursive Kalman Filter (R3SRKF) for IEEE~1588 clock tracking under asymmetric delays~\cite{10649634}. The estimator predicts the clock state, then estimates the asymmetry-induced bias, and finally corrects the clock state using the bias-compensated observation.

\paragraph{Clock-State Prediction}
Before incorporating the new timestamp observation, the clock state is first propagated from the previous synchronization round. Given $\boldsymbol{\Phi}_k$ and $\boldsymbol{Q}_{d,k}$, the prior clock state and covariance are computed as
\begin{align} 
\hat{\boldsymbol{x}}_{k|k-1} &= \boldsymbol{\Phi}_k\hat{\boldsymbol{x}}_{k-1|k-1},\\ \boldsymbol{P}_{k|k-1} &= \boldsymbol{\Phi}_k \boldsymbol{P}_{k-1|k-1} \boldsymbol{\Phi}_k^\top + \boldsymbol{Q}_{d,k}, 
\label{eq:r3srkf-pred-cov} 
\end{align} 
where $\boldsymbol{P}_{k|k-1}\in\mathbb{R}^{2\times 2}$ is the predicted
covariance of the clock state $\boldsymbol{x}_k=[\theta_k,\varpi_k]^\top$,
representing the uncertainty of the prior offset and skew estimates.

After prediction, the timestamp observation is compared with the predicted clock state. The observation residual can be decomposed as
\begin{equation}
\boldsymbol{r}_k
=
\boldsymbol{z}_k
-
\boldsymbol{A}\hat{\boldsymbol{x}}_{k|k-1}
=
\boldsymbol{A}
\left(
\boldsymbol{x}_k-\hat{\boldsymbol{x}}_{k|k-1}
\right)
+
\boldsymbol{H}b_k
+
\boldsymbol{\eta}_k,
\label{eq:r3srkf-residual-decomp}
\end{equation}
which contains the clock-state prediction error, the asymmetry-induced bias, and the timestamp-observation noise. Since the asymmetry-induced bias is estimated separately, the residual covariance used for bias estimation is formed from the clock-state prediction uncertainty and the observation noise:
\begin{equation}
\boldsymbol{G}_{r,k}
=
\boldsymbol{A}
\boldsymbol{P}_{k|k-1}
\boldsymbol{A}^\top
+
\boldsymbol{I}_k .
\label{eq:random-residual-cov}
\end{equation}

\paragraph{Asymmetry-Bias Estimation}
Based on $\boldsymbol{G}_{r,k}$, the R3SRKF estimates the asymmetry-induced bias contained in the residual before correcting the clock state. The corresponding bias-estimation gain is given by
\begin{equation}
\boldsymbol{K}_{b,k}
=\left(\boldsymbol{H}^\top
\boldsymbol{G}_{r,k}^{-1}
\boldsymbol{H}
\right)^{-1}
\boldsymbol{H}^\top
\boldsymbol{G}_{r,k}^{-1}.
\label{eq:bias-gain}
\end{equation}

The asymmetry-induced bias is then estimated from the residual as
\begin{equation}
\hat{b}_k
=
\boldsymbol{K}_{b,k}\boldsymbol{r}_k,
\label{eq:bias-est}
\end{equation}
with covariance $
S_{b,k}
=
\left(
\boldsymbol{H}^\top
\boldsymbol{G}_{r,k}^{-1}
\boldsymbol{H}\right)^{-1}$.  

\paragraph{Clock-State Correction}
{After estimating the asymmetry-induced bias, its contribution is removed from the observation residual. The bias-compensated residual is thus:} 
\begin{equation} 
\tilde{\boldsymbol{r}}_k = \boldsymbol{z}_k - \boldsymbol{A} \hat{\boldsymbol{x}}_{k|k-1} - \boldsymbol{H}\hat{b}_k . \label{eq:bias-compensated-residual} 
\end{equation} 

The predicted clock state is then corrected using the bias-compensated residual:
\begin{equation} 
\hat{\boldsymbol{x}}_{k|k} = \hat{\boldsymbol{x}}_{k|k-1} + \boldsymbol{P}_{k|k-1} \boldsymbol{A}^\top \boldsymbol{G}_{r,k}^{-1} \tilde{\boldsymbol{r}}_k, 
\label{eq:r3srkf-state-update} 
\end{equation}

Subsequently, the posterior covariance is updated 
\begin{align} 
\boldsymbol{P}_{k|k} &= \boldsymbol{P}_{k|k-1} \\ 
&- \boldsymbol{P}_{k|k-1} \boldsymbol{A}^\top \boldsymbol{G}_{r,k}^{-1} \left( \boldsymbol{G}_{r,k} - \boldsymbol{H}S_{b,k}\boldsymbol{H}^\top \right) \boldsymbol{G}_{r,k}^{-1} \boldsymbol{A} \boldsymbol{P}_{k|k-1}. \notag
\label{eq:r3srkf-cov-update} 
\end{align}

In this clock-state model, the asymmetry-induced bias $b_k$ is derived from the current observation residual and used to compensate the timestamp observation, while the recursively propagated clock state remains $\boldsymbol{x}_k=[\theta_k,\varpi_k]^\top$. Consider two timestamp observations $t_k$ and $t_{k+1}$. Taking $t_k$ as the reference time, the offset evolution is approximated as $ \theta(t_{k+1}) \approx \theta_k+\varpi_k(t_{k+1}-t_k)$, 
the local observability matrix for $\boldsymbol{x}_k=[\theta_k,\varpi_k]^\top$ is  thus
\begin{equation} 
\mathcal{O}_k = 
\begin{bmatrix} 
1 & 0\\ 1 & \Delta t_{k+1} 
\end{bmatrix},
\label{eq:observability-matrix} 
\end{equation} 
where $\Delta t_{k+1}=t_{k+1}-t_k$. Since  the determinant  $\det(\mathcal{O}_k)=\Delta t_{k+1}$,  we consider the clock offset and relative skew are locally observable when $\Delta t_{k+1}\neq 0$.

Overall, the three-step recursive estimator provides the corrected clock-state estimate $(\hat{\theta}_{k|k},\hat{\varpi}_{k|k})$ and the covariance terms $S_{b,k}$ and $\boldsymbol{P}_{k|k}$. The corrected clock state is used for synchronized timestamp mapping, while the covariance terms are propagated to the subsequent reliability assessment.

\subsubsection{Synchronized Timestamp Mapping and AoI Definition}
\label{sec:freshness-metrics}

To evaluate feature freshness on a consistent temporal reference, we first define the synchronized timestamp mapping and the relevant AoI metrics for feature alignment and fusion.

\begin{definition}[Synchronized Timestamp Mapping]
\label{def_1}
For ego vehicle $m$ and neighboring vehicle $l\in{L}_m$, let $t_l$
denote a local timestamp recorded by vehicle $l$. To express this timestamp on
the ego temporal reference, we compensate for the estimated clock offset and
relative skew after asymmetry-bias correction:
\begin{equation}
\mathrm{Sync}_{l,m}(t_l)
\triangleq
t_l
-
\hat{\theta}_{lm,k|k}
-
\hat{\varpi}_{lm,k|k}(t_l-t_k),
\label{eq:sync_mapping}
\end{equation}
where $t_k$ is the latest synchronization exchange time. This mapping converts vehicle-local timestamps into the ego temporal reference, enabling consistent freshness evaluation and temporal feature compensation.
\end{definition}

\begin{definition}[Arrival AoI]
\label{def_3}
At ego-reference time $t$, let $g_{lm}^{\mathrm{arr}}(t)$ be the local
generation timestamp of the latest feature update from vehicle $l$ available at vehicle $m$. According to Definition~\ref{def_1}, its synchronized generation time on the ego temporal reference is
\begin{equation}
u_{lm}(t)
\triangleq
\mathrm{Sync}_{l,m}\big(g_{lm}^{\mathrm{arr}}(t)\big).
\label{eq:arrival_generation_time}
\end{equation}

The arrival AoI is defined as
\begin{equation}
\mathcal{A}_{lm}(t)
\triangleq
t-u_{lm}(t),
\label{eq:arrival_aoi}
\end{equation}
which characterizes the age of the latest feature update from vehicle $l$ that
is already available at vehicle $m$.
\end{definition}

\begin{figure}[!t]
\centering
\includegraphics[width=2.8in]{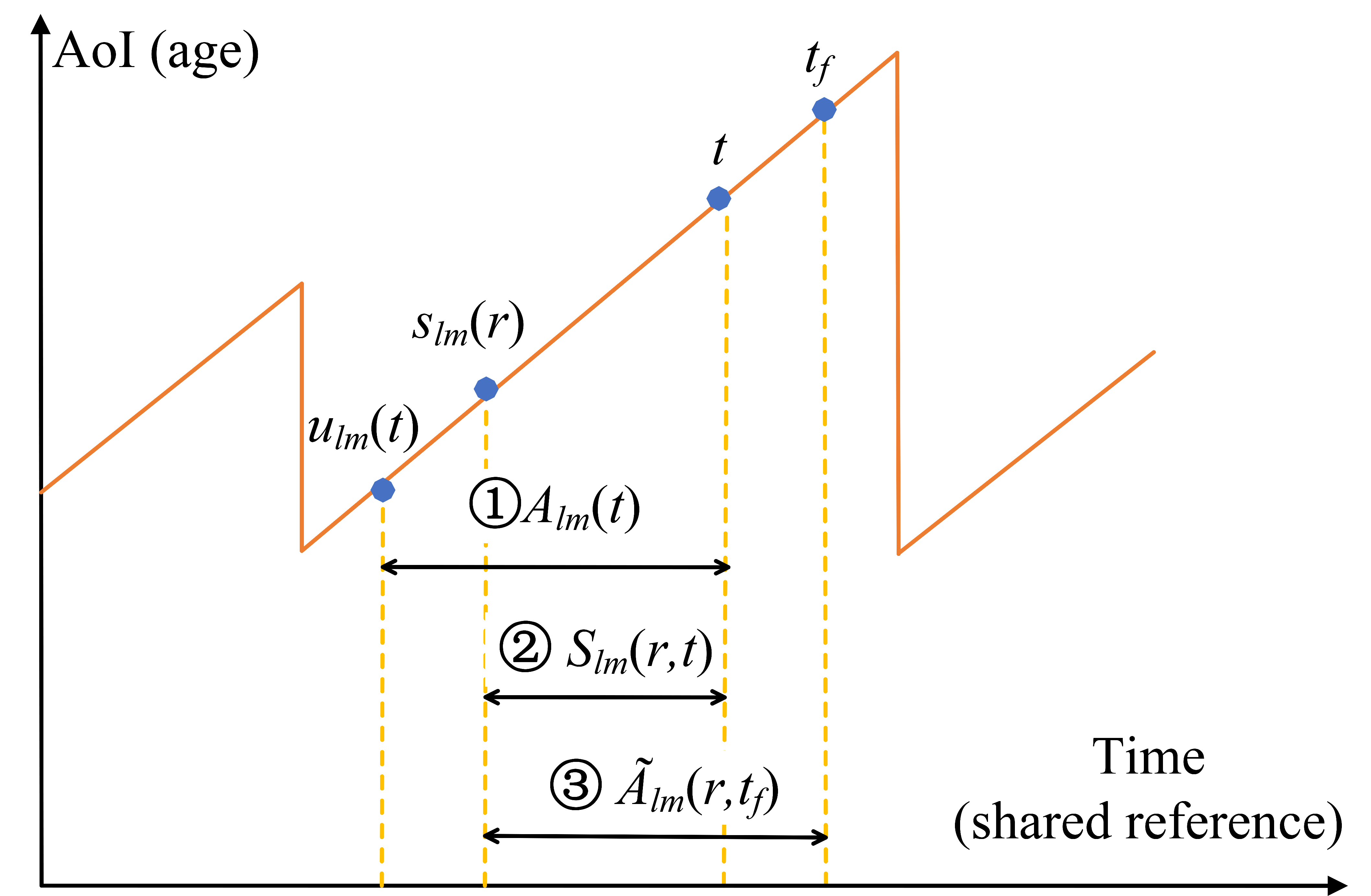}
\caption{Freshness metrics on a common temporal reference:
\ding{172} arrival AoI $\mathcal{A}_{lm}(t)$ for the latest
arrived feature update;
\ding{173} current age $\mathcal{S}_{lm}(r,t)$ of candidate RoI $r$;
\ding{174} delivery-time AoI $\tilde{\mathcal{A}}_{lm}(r,t_f)$, which
accounts for the expected additional aging during inter-vehicle
communication before the feature becomes available for fusion.}
\label{fig_3}
\end{figure}

\begin{proposition}
The arrival AoI $\mathcal{A}_{lm}(t)$ in Definition~\ref{def_3} follows the standard sawtooth evolution:
\begin{itemize}
\item When a new feature update arrives at vehicle $m$, $u_{lm}(t)$ is updated to the synchronized generation time of the newly arrived feature;
\item Between two consecutive arrivals, $u_{lm}(t)$ remains unchanged, and $\mathcal{A}_{lm}(t)$ increases with unit slope, i.e.,
$\frac{d}{dt}\mathcal{A}_{lm}(t)=1$.
\end{itemize}

As illustrated in Fig.~\ref{fig_3}, the arrival AoI drops at each feature arrival and increases linearly between arrivals.
\end{proposition}

The arrival AoI $\mathcal{A}_{lm}(t)$ follows the conventional AoI definition
and characterizes the receiver-side freshness of feature updates that are
already available at ego vehicle $m$~\cite{yates2021aoi_survey}. It does not directly describe a candidate RoI whose feature payload has not yet become available for fusion. In collaborative feature sharing, RoI selection and weighting are performed for such candidate RoIs before their payloads reach the
ego vehicle. Therefore, the freshness of a candidate RoI should be evaluated
using its own synchronized generation time and the additional aging during
inter-vehicle communication. We define delivery-time AoI to quantify the
expected age of the candidate RoI feature when it becomes available for fusion.

\begin{definition}[Delivery-Time AoI]
\label{def_4}
For a candidate RoI $r$ generated by vehicle $l$, let
$g_l^{\mathrm{feat}}(r)$ denote the local generation timestamp of its
corresponding feature. According to Definition~\ref{def_1}, its synchronized
generation time on the ego temporal reference is
\begin{equation}
s_{lm}(r)
\triangleq
\mathrm{Sync}_{l,m}\big(g_l^{\mathrm{feat}}(r)\big).
\label{eq:candidate_generation_time}
\end{equation}

At time $t$, the current age of this candidate RoI feature is
\begin{equation}
\mathcal{S}_{lm}(r,t)
\triangleq
t-s_{lm}(r).
\label{eq:candidate_feature_age}
\end{equation}

Let $\widehat{\Delta t}^{\mathrm{comm}}_{lm}(r,t)$ denote the expected
communication delay from time $t$ until the corresponding feature becomes
available at vehicle $m$ for fusion. The delivery-time AoI is defined as
\begin{equation}
\tilde{\mathcal{A}}_{lm}(r,t_f)
\triangleq
\mathcal{S}_{lm}(r,t)
+
\widehat{\Delta t}^{\mathrm{comm}}_{lm}(r,t),
\label{eq:delivery_aoi}
\end{equation}
where $t_f$ denotes the expected fusion time determined by the
communication delay. Thus, $\tilde{\mathcal{A}}_{lm}(r,t_f)$ estimates the age of RoI $r$ when its feature becomes available for fusion.
\end{definition}

We next present an example to indicate these definitions.

\begin{example}
\label{ex:ages}
Consider ego vehicle $m$ and neighboring vehicle $l$. The latest synchronization exchange occurs at $t_k=10~\mathrm{s}$, and the RoI selection and weighting time is $t=10.25~\mathrm{s}$ on the ego temporal reference. After asymmetry-bias correction, the estimated relative clock state is $\hat{\theta}_{lm,k|k}=-10~\mathrm{ms}$ and $\hat{\varpi}_{lm,k|k}=10~\mathrm{ppm}$.

\textbf{(1) Current age of candidate RoI $\mathcal{S}_{lm}(r,t)$:}
Assume that the feature associated with candidate RoI $r$ is generated at
vehicle $l$ with local timestamp $g_l^{\mathrm{feat}}(r)=10.20~\mathrm{s}$.
According to Definition~\ref{def_1}, its synchronized generation time on the
ego temporal reference is
\begin{align*}
s_{lm}(r)
&=
\mathrm{Sync}_{l,m}
\big(g_l^{\mathrm{feat}}(r)\big) \\
&=
10.20
-
(-0.010)
-
(10^{-5})(10.20-10)
\approx
10.21~\mathrm{s}.
\end{align*}

Thus, the current age of this candidate RoI is
\begin{align*}
\mathcal{S}_{lm}(r,t)
=	t-s_{lm}(r) =
10.25-10.21
=
0.040~\mathrm{s}
=
40~\mathrm{ms}.
\end{align*}

\textbf{(2) Arrival AoI $\mathcal{A}_{lm}(t)$:}
By time $t$, suppose the latest feature update from vehicle $l$ that has
already arrived at vehicle $m$ was generated at local time
$g_{lm}^{\mathrm{arr}}(t)=10.16~\mathrm{s}$. Its synchronized generation time is
\begin{align*}
u_{lm}(t)
&=
\mathrm{Sync}_{l,m}
\big(g_{lm}^{\mathrm{arr}}(t)\big) \\
&=
10.16
-
(-0.010)
-
(10^{-5})(10.16-10)
\approx
10.17~\mathrm{s}.
\end{align*}

Therefore, the arrival AoI at time $t$ is
\begin{align*}
\mathcal{A}_{lm}(t)
=
t-u_{lm}(t) =
10.25-10.17
=
0.080~\mathrm{s}
=
80~\mathrm{ms}.
\end{align*}

\textbf{(3) Delivery-time AoI $\tilde{\mathcal{A}}_{lm}(r,t_f)$:}
For candidate RoI $r$, suppose the expected communication delay from time $t$ to feature availability is
$\widehat{\Delta t}^{\mathrm{comm}}_{lm}(r,t)=25~\mathrm{ms}$. The corresponding fusion time is
\begin{align*}
t_f
=
t+\widehat{\Delta t}^{\mathrm{comm}}_{lm}(r,t) 
=
10.25+0.025
=
10.275~\mathrm{s}.
\end{align*}

By Definition~\ref{def_4}, the delivery-time AoI is
\begin{align*}
\tilde{\mathcal{A}}_{lm}(r,t_f)
&=
\mathcal{S}_{lm}(r,t)
+
\widehat{\Delta t}^{\mathrm{comm}}_{lm}(r,t) \\
&=
40~\mathrm{ms}
+
25~\mathrm{ms}
=
65~\mathrm{ms}.
\end{align*}
\end{example}
\subsubsection{Synchronization Uncertainty and Reliability}
\label{sec:temporal-uncertainty}

The synchronized timestamp mapping in Definition~\ref{def_1} and the
delivery-time AoI in Definition~\ref{def_4} rely on the corrected relative clock-state estimate $(\hat{\theta}_{lm,k|k},\hat{\varpi}_{lm,k|k})$.   {Since this estimate is obtained from timestamp exchanges, sparse synchronization updates, timestamp perturbations, and forward--reverse delay asymmetry can leave residual timing uncertainty. Such uncertainty propagates through the timestamp mapping and affects the temporal reliability of shared features. Therefore, we quantify the synchronization uncertainty and convert it into a reliability factor for feature fusion.}

To quantify this residual uncertainty, we use the posterior covariance
$\boldsymbol{P}_{k|k}$ produced by the recursive estimator, which describes the uncertainty of the corrected clock state. We write
\begin{equation} 
\boldsymbol{P}_{k|k} = 
\begin{bmatrix} P_{\theta\theta} & P_{\theta\varpi}\\ P_{\theta\varpi} & P_{\varpi\varpi} 
\end{bmatrix}, 
\end{equation} 
where $P_{\theta\theta}$ is the offset variance, $P_{\varpi\varpi}$ is the skew variance, and $P_{\theta\varpi}$ is their covariance.

After the latest clock-state update, the corrected offset and skew estimates are propagated over the elapsed interval $\Delta t$ when a feature timestamp is mapped onto the ego temporal reference. According to the clock-state transition model in Eqs.~\eqref{eq:clock-state-transition}--\eqref{eq:Qd-k-revised}, the
residual synchronization uncertainty is
\begin{align}
\label{eq:theta-growth-k}
\sigma_{\mathrm{sync}}^2(\Delta t)
=
& P_{\theta\theta}
+2\Delta t P_{\theta\varpi}
+\Delta t^2 P_{\varpi\varpi}
+q_\theta\Delta t \notag\\
&+\frac{1}{3}q_\varpi\Delta t^3
+S_{b,k}.
\end{align}

In Eq.~\eqref{eq:theta-growth-k}, $P_{\theta\theta}$ represents the residual
offset uncertainty after the latest clock-state update and contributes a timing
error that is approximately independent of $\Delta t$. The terms
$2\Delta t P_{\theta\varpi}$ and $\Delta t^2 P_{\varpi\varpi}$ describe how
the offset--skew correlation and residual skew uncertainty propagate as the
elapsed interval from the latest synchronization update increases. Thus,
residual offset mainly causes a nearly fixed timestamp bias, whereas residual
skew causes the timestamp uncertainty to increase with the elapsed interval.
The terms $q_\theta\Delta t$ and $\frac{1}{3}q_\varpi\Delta t^3$ further account
for uncertainty growth caused by clock process noise, while $S_{b,k}$ accounts
for the residual uncertainty after asymmetry-bias correction. This timing
uncertainty affects both delivery-time AoI and spatiotemporal feature alignment.
A residual timing error changes the estimated temporal gap used for feature
compensation and, under relative vehicle motion, can induce BEV feature
misalignment. Therefore, the resulting
$\sigma_{\mathrm{sync}}^2(\Delta t)$ is used to evaluate the temporal
reliability of shared features.

\begin{figure}[!t]
\centering
\includegraphics[width=0.7\columnwidth]{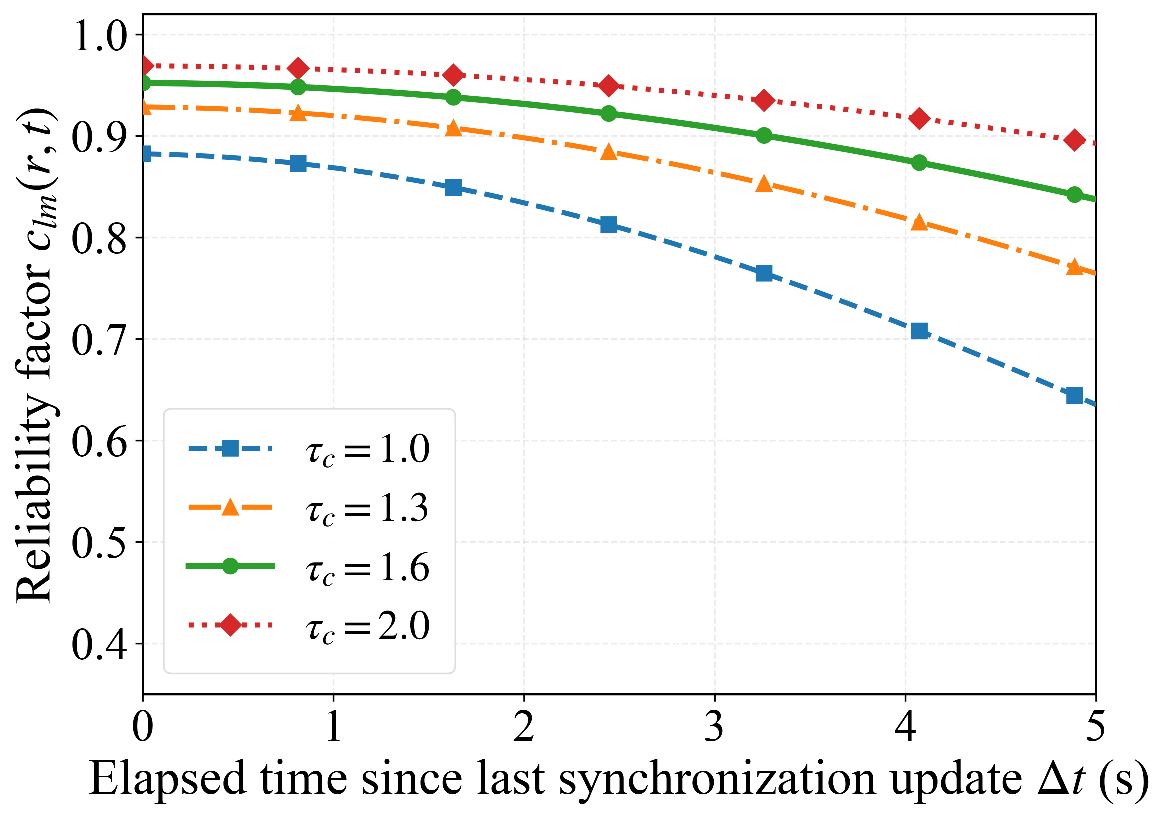}
\caption{ {Sensitivity of the synchronization reliability to $\tau_c$.}}
\label{fig:tau_c_sensitivity}
\end{figure}

For RoI $r$ evaluated at time $t$, let $\Delta t_{lm}^{\mathrm{sync}}(r,t)$ be the elapsed interval since the latest synchronization update between vehicles $l$ and $m$. Substituting this interval into Eq.~\eqref{eq:theta-growth-k} gives the residual synchronization uncertainty used for timestamp mapping. We define the corresponding reliability factor as
\begin{equation}
c_{lm}(r,t)
=
\exp\left(
-
\frac{
\sigma_{\mathrm{sync}}^2
\left(
\Delta t_{lm}^{\mathrm{sync}}(r,t)
\right)
}
{\tau_c^2}
\right),
\label{eq:synchronization_reliability}
\end{equation}
where $\tau_c>0$ controls the sensitivity of $c_{lm}(r,t)$ to residual synchronization uncertainty. A smaller $\tau_c$ makes the reliability factor decay faster as the synchronization uncertainty increases, while a larger $\tau_c$ weakens this penalty. Fig.~\ref{fig:tau_c_sensitivity} illustrates the
effect of $\tau_c$ on the reliability factor as the elapsed time since the latest synchronization update increases. In our implementation, $\tau_c=1.6$.

{\subsubsection{Synchronization Overhead}
The synchronization overhead mainly comes from the periodic timestamp exchanges used to maintain the relative clock states. Let $B_{\mathrm{sync}}$   denote the number of bits required by one
timestamp-exchange round  between vehicle $m$ and neighbor $l\in{L}_m$, $T_{\mathrm{sync}}$ denote the synchronization period. For vehicle $m$, the timestamp-exchange traffic over $K$ exchange rounds is $C_m^{\mathrm{sync}}=K|{L}_m|B_{\mathrm{sync}}$, and the  average timestamp-exchange rate is $R_m^{\mathrm{sync}}
=|{L}_m|B_{\mathrm{sync}}/T_{\mathrm{sync}}$. Thus, the synchronization overhead increases with the number of maintained relative clock states and decreases with a larger synchronization period.}

Aggregating the timestamp-exchange traffic over all vehicles gives
$R_{\mathrm{net}}^{\mathrm{sync}}={B_{\mathrm{sync}}}/{T_{\mathrm{sync}}}\sum_{m=1}^{M}|{L}_m|$.
With the average neighbor degree $\bar d={1}/{M}\sum_{m=1}^{M}|{L}_m|$,
the above expression becomes $R_{\mathrm{net}}^{\mathrm{sync}}=
M\bar d B_{\mathrm{sync}}/T_{\mathrm{sync}}$. 
Therefore, the synchronization overhead scales linearly with the number of vehicles when $\bar d$ is bounded, while the worst-case $O(M^2)$ scaling occurs  when all neighboring vehicles  maintain  synchronization with the ego vehicle.

\subsection{Communication Model}
\label{subsec:comm_model}

Based on the constructed temporal reference, the freshness of a candidate RoI should be evaluated when it becomes available for fusion, depending on the feature-generation timestamp and the communication delay required to deliver the RoI payload. Therefore, this subsection models the communication delay used in delivery-time AoI calculation.

In collaborative perception, feature sharing mainly occurs among nearby vehicles and satisfies the latency and reliability requirements of V2X services~\cite{3gpp_ts22186}. Accordingly, we model inter-vehicle feature delivery using New Radio (NR)-V2X sidelink communication over the PC5 interface, which supports direct vehicle-to-vehicle information exchange. Under Mode~2 operation, vehicles autonomously select resources from a configured or pre-configured sidelink resource pool~\cite{3gpp_tr37885,ali2021nr_v2x_mode2}. This resource-selection process introduces access latency before the RoI payload can be transmitted. 

After resource access, RoI transmission depends on the payload size, occupied subchannels, channel quality, and selected modulation and coding scheme (MCS). We capture these factors through a sidelink link-abstraction model over an Orthogonal Frequency-Division Multiplexing (OFDM)-based resource grid~\cite{3gpp_ts38211,3gpp_ts38214}. In this model, signal-to-interference-plus-noise ratios (SINRs) are mapped to packet decoding reliability through effective-SINR and MCS-dependent SINR--block-error-rate (BLER) curves~\cite{lagen2020nr_phy_abstraction,lusvarghi2024link_level}.
Together, the access latency and RoI-payload transmission delay determine when the shared RoI becomes available for fusion.

\subsubsection{Sidelink Channel Model}
\label{subsubsec:sidelink_channel}
Consider a sidelink resource pool partitioned into a set of subchannels $\mathcal{E}=\{1,2,\ldots,N_e\}$, where each subchannel has bandwidth $W_{\mathrm{sub}}$. For a candidate RoI transmitted from vehicle $l$ to vehicle $m$, the SINR on subchannel $e$ is modeled as
\begin{equation} 
\gamma_{l,m}^{e} = \frac{P_l(e)h_{l,m}^{e}} {\displaystyle \sum\nolimits_{j\in\mathcal{I}_{m,e}} P_j(e)h_{j,m}^{e} + N_0 W_{\mathrm{sub}}}, 
\label{eq:sinr_comm} 
\end{equation} 
where $P_l(e)$ is the transmit power of vehicle $l$ on subchannel $e$, $h_{l,m}^{e}$ is the channel gain from $l$ to $m$, $\mathcal{I}_{m,e}$ is the set of co-channel interferers at vehicle $m$ (receiver) on subchannel $e$, and $N_0$ is the noise spectral density. Eq.~\eqref{eq:sinr_comm} provides the sidelink quality used by the link-abstraction model. When no co-channel interferer occupies subchannel $e$, $\mathcal{I}_{m,e}=\varnothing$, and Eq.~\eqref{eq:sinr_comm} reduces to the corresponding SNR expression.

\subsubsection{RoI Payload Size and Subchannel Occupation}
\label{subsubsec:roi_payload_subchannel}

We next relate the RoI feature payload to the sidelink resource occupation. The BEV feature map is divided into an $N_G\times N_G$ RoI grid, and let
$\mathcal{R}=\{1,\ldots,R\}$ be the set of RoI indices, $R=N_G^2$.
Each RoI $r\in\mathcal{R}$ covers a set of BEV feature cells denoted by
$\Omega_r$. These RoIs form a non-overlapping partition of the BEV feature map,
i.e., $\bigcup_{r=1}^{R}\Omega_r=\Omega_{\mathrm{BEV}}$, with $\Omega_i\cap\Omega_j=\emptyset$ and $i\neq j$.

For RoI $r$, let $N_r^{\mathrm{grid}}=|\Omega_r|$ be the number of BEV feature
cells covered by this RoI. Since each BEV feature cell contains
$C_{\mathrm{feat}}$ feature channels and each channel entry is represented by
$q$ bits, the feature payload size of RoI $r$ is given by
\begin{equation}
B(q,r)
=
B_{\mathrm{hdr}}
+
q C_{\mathrm{feat}} N_r^{\mathrm{grid}},
\label{eq:bit_cost}
\end{equation}
where $B_{\mathrm{hdr}}$ denotes the packet header and metadata overhead.

Given the selected MCS $\chi_{lm}$ for the sidelink transmission from vehicle $l$ to vehicle $m$, the payload capacity of one subchannel is denoted by $C_{\mathrm{sub}}(\chi_{lm})$. This capacity is determined by the configured subchannel resource size and the spectral efficiency of the selected MCS. Since sidelink resources are assigned in subchannel units, the number of subchannels required for RoI $r$ is
\begin{equation}
n_{lm}^{\mathrm{sub}}(r)
=
\left\lceil
\frac{B(q,r)}
{C_{\mathrm{sub}}(\chi_{lm})}
\right\rceil,
\label{eq:subchannel_occupation}
\end{equation}
which gives the subchannel occupation used in the following link-abstraction
model~\cite{3gpp_ts38214,yan2022mcs_v2x,lusvarghi2024link_level}.

\subsubsection{MCS-Dependent Sidelink Link Abstraction}
\label{subsubsec:mcs_link_abstraction}

Based on the RoI payload size and subchannel occupation, we evaluate the decoding
reliability of each RoI transmission using a sidelink link-abstraction model.
The model maps the  SINRs over the occupied subchannels into an effective SINR, and then obtains the physical sidelink shared channel (PSSCH) BLER under the selected MCS~\cite{lagen2020nr_phy_abstraction,lusvarghi2024link_level}.

Specifically, let $\mathcal{E}_{lm}(r)\subseteq\mathcal{E}$ denote the set of
subchannels used for transmitting RoI $r$ from vehicle $l$ to vehicle $m$, with
$|\mathcal{E}_{lm}(r)|=n_{lm}^{\mathrm{sub}}(r)$. The effective SINR of this
RoI transmission is
\begin{equation}
\bar{\gamma}_{lm}(r)
=
\Phi_{\mathrm{SINR}}
\left(
\left\{\gamma_{lm}^{e}\right\}_{e\in\mathcal{E}_{lm}(r)}
\right),
\label{eq:effective_sinr}
\end{equation}
where $\Phi_{\mathrm{SINR}}(\cdot)$ is the effective-SINR mapping used in link-to-system abstraction. It compresses the SINRs over the occupied subchannels into a scalar $\bar{\gamma}_{lm}(r)$, which is used as the input to the MCS-dependent SINR--BLER curve. The mapping can be implemented using Exponential Effective SINR Mapping (EESM), or using the effective-SINR calculation adopted by the simulator to fit the BLER model~\cite{lagen2020nr_phy_abstraction,patriciello2019e2e}.

The PSSCH BLER of the RoI packet is represented by
\begin{equation}
\Lambda_{lm}^{\mathrm{SL}}(r)
=
\Lambda_{\mathrm{PSSCH}}^{\mathrm{SL}}
\left(
\bar{\gamma}_{lm}(r),
\chi_{lm},
B(q,r),
n_{lm}^{\mathrm{sub}}(r)
\right),
\label{eq:sidelink_bler}
\end{equation}
where  $\Lambda_{\mathrm{PSSCH}}^{\mathrm{SL}}(\cdot)$ denotes the sidelink PSSCH BLER model, and can be represented by an SINR--BLER curve or lookup table~\cite{lagen2020nr_phy_abstraction,lusvarghi2024link_level}. In our implementation, $\Lambda_{\mathrm{PSSCH}}^{\mathrm{SL}}(\cdot)$ is instantiated using a logistic SINR--BLER curve fitted from ns-3 5G-LENA NR-V2X packet reception records.

Based on the PSSCH BLER  in Eq.~\eqref{eq:sidelink_bler}, we define the effective payload throughput for RoI transmission as
\begin{equation}
R_{lm}^{\mathrm{eff}}(r)
=
n_{lm}^{\mathrm{sub}}(r) W_{\mathrm{sub}}
\zeta(\chi_{lm})
\left[1-\Lambda_{lm}^{\mathrm{SL}}(r)\right],
\label{eq:effective_throughput_sl}
\end{equation}
where $\zeta(\chi_{lm})$ is the spectral efficiency of the selected MCS, and $1-\Lambda_{lm}^{\mathrm{SL}}(r)$ is the successful decoding probability of the RoI packet. Thus, $R_{lm}^{\mathrm{eff}}(r)$ captures the MCS-dependent trade-off between spectral efficiency and decoding reliability. A higher MCS can increase the payload throughput, but may also increase the BLER when the effective SINR is insufficient~\cite{yan2022mcs_v2x, lusvarghi2024link_level}.

Based on the above payload size $B(q,r)$ and effective payload throughput  $R_{lm}^{\mathrm{eff}}(r)$, the RoI-payload transmission delay is calculated as
\begin{equation}
\Delta t^{\mathrm{tx}}_{lm}(r)
=
\frac{B(q,r)}
{R_{lm}^{\mathrm{eff}}(r)}.
\label{eq:rate_based_delivery_time}
\end{equation}

\subsubsection{Sidelink Access Latency and Communication Delay}
\label{subsubsec:comm_delay}
{
Under NR-V2X Mode~2, a vehicle selects sidelink resources autonomously from a configured or pre-configured resource pool before transmitting the RoI payload~\cite{3gpp_tr37885,ali2021nr_v2x_mode2}. This resource-selection process introduces access latency.
}

{
The resource selection is performed within a selection window bounded by
$T_1$ and $T_2$~\cite{3gpp_tr37885,ali2021nr_v2x_mode2,5gaa_whitepaper_2024},
where $T_1$ and $T_2$ denote the lower and upper bounds of the candidate
resource window. The upper bound $T_2$ is associated with the packet delay budget and latency requirement, while the realized access latency also depends on the resource-pool condition, such as channel congestion and resource contention~\cite{etsi_tr103766}. We model the access latency from vehicle $l$ to vehicle $m$ as
\begin{equation}
\Delta t^{\mathrm{acc}}_{lm}
=
\mathcal{T}_{\mathrm{acc}}
\bigl(T_1,T_2,\kappa_{lm}\bigr),
\label{eq:access_latency}
\end{equation}
where $\kappa_{lm}$ denotes the sidelink resource-pool condition, and
$\mathcal{T}_{\mathrm{acc}}(\cdot)$ is instantiated with condition-specific access-latency settings obtained from NR-V2X sidelink simulation.
}

{
Overall, the expected communication delay is given by
\begin{equation}
\widehat{\Delta t}^{\mathrm{comm}}_{lm}(r,t)
=
\Delta t^{\mathrm{acc}}_{lm}(t)
+
\Delta t^{\mathrm{tx}}_{lm}(r,t).
\label{eq:comm_delay_total}
\end{equation}}

{Therefore, the candidate RoI is expected to
become available at vehicle $m$ after $\widehat{\Delta t}^{\mathrm{comm}}_{lm}(r,t)$, and its delivery-time AoI is	computed as in Definition~\ref{def_4}.}

\subsection{Spatiotemporal Feature Alignment}
\label{sec:spatial-align}

Considering that neighbor and ego features are often generated at different timestamps and expressed in different coordinate frames, they cannot be directly fused. Therefore, we design spatiotemporal feature alignment, where geometric projection first maps the neighbor feature, generated at its synchronized timestamp, into the ego coordinate frame at $t_f$, and temporal feature compensation adjusts the projected feature to reduce the temporal mismatch before fusion.

\subsubsection{Geometric Projection}

For a candidate RoI $r$ generated by vehicle $l$, its synchronized generation time on the ego temporal reference is $s_{lm}(r)$, according to Eq.~\eqref{eq:candidate_generation_time}. The corresponding feature is associated with coordinate frame $\mathcal{F}_l(s_{lm}(r))$, while the ego feature used for fusion is represented in $\mathcal{F}_m(t_f)$. We define the rigid transformation from $\mathcal{F}_l(s_{lm}(r))$ to $\mathcal{F}_m(t_f)$ as
\begin{equation}
\mathcal{T}_{l\rightarrow m}(r,t_f)
=
\left(
\boldsymbol{R}_{lm}(t_f,s_{lm}(r)),
\boldsymbol{n}_{lm}(t_f,s_{lm}(r))
\right),
\end{equation}
where $\boldsymbol{R}_{lm}(t_f,s_{lm}(r))$ and
$\boldsymbol{n}_{lm}(t_f,s_{lm}(r))$ are the relative rotation and translation
computed from the poses of vehicles $m$ and $l$ at the corresponding
timestamps. For a BEV location $\boldsymbol{p}_l$ in the neighbor frame, its projected coordinate in the ego fusion frame is
\begin{equation}
\boldsymbol{p}_m
=
\mathcal{T}_{l\rightarrow m}(r,t_f)\boldsymbol{p}_l
=
\boldsymbol{R}_{lm}(t_f,s_{lm}(r))\boldsymbol{p}_l
+
\boldsymbol{n}_{lm}(t_f,s_{lm}(r)).
\label{eq:rigid-transform}
\end{equation}

Applying this transformation to the retained RoI features yields the projected neighbor feature $\tilde{\boldsymbol{F}}_l(t_f)\in\mathbb{R}^{X'\times Y'\times C}$ in the ego fusion frame. This projection resolves the coordinate-frame difference between the neighbor feature and the ego feature $\boldsymbol{F}_m(t_f)$. 

\subsubsection{Temporal Feature Compensation}
Although geometric projection places the neighbor feature in the ego coordinate frame, it does not remove the temporal mismatch between the neighbor observation and the ego fusion time. Specifically, the projected feature of RoI $r$ still corresponds to the scene at its synchronized generation time $s_{lm}(r)$, whereas fusion is performed at $t_f$. We therefore introduce a temporal feature compensation module  to adjust the projected neighbor feature before fusion.

Using the RoI partition defined in Sec.~\ref{subsubsec:roi_payload_subchannel},
let $\mathcal{S}_{lm}(r,t_f)$ be the age of RoI $r$ at the fusion time $t_f$, measured on the ego temporal reference. We generate a BEV-grid age map by
assigning this RoI-level age to every BEV location within the corresponding RoI, i.e., $\mathcal{S}_{lm}(\boldsymbol{\upsilon},t_f)=\mathcal{S}_{lm}(r,t_f), \boldsymbol{\upsilon}\in\Omega_r$,  
The resulting BEV-grid age map is encoded by a two-layer multilayer perceptron (MLP):
\begin{equation}
\boldsymbol{a}_{lm}(\boldsymbol{\upsilon},t_f)
=
\phi_{\mathrm{age}}
\left(
\mathcal{S}_{lm}(\boldsymbol{\upsilon},t_f)
\right),
\label{eq:age-embed}
\end{equation}
where $\phi_{\mathrm{age}}(\cdot)$ maps the feature age into a temporal
embedding. This embedding represents the staleness of the neighbor feature at
each BEV location before fusion.

{To compensate the time-induced feature shift, the projected neighbor feature
and the temporal  embedding are concatenated and fed into a Conv2D-based flow head:
\begin{equation}
	\boldsymbol{\vartheta}_{l}(t_f)
	=
	{\cal B}_{\max}
	\tanh
	\left(
	\mathcal{H}
	\left(
	[
	\tilde{\boldsymbol{F}}_{l}(t_f);
	\boldsymbol{a}_{lm}(t_f)
	]
	\right)
	\right),
	\label{eq:flow-predict}
\end{equation}
where $[\cdot;\cdot]$ denotes channel-wise concatenation. $\mathcal{H}(\cdot)$ is a lightweight convolutional encoder--decoder that outputs a two-channel BEV displacement field. The two channels represent the horizontal and vertical sampling offsets at each BEV location. $\tanh(\cdot)$ bounds the normalized displacement, and ${\cal B}_{\max}$ scales it to the maximum allowed BEV displacement.}

{
The projected feature is then warped according to the predicted displacement:
\begin{equation}
	\bar{\boldsymbol{F}}_{l}(t_f,\boldsymbol{\upsilon})
	=
	\tilde{\boldsymbol{F}}_{l}
	\left(
	t_f,
	\boldsymbol{\upsilon}
	+
	\boldsymbol{\vartheta}_{l}(t_f,\boldsymbol{\upsilon})
	\right),
	\label{eq:flow-warp}
\end{equation}
where bilinear sampling is used. In this way, the projected neighbor feature is
temporally adjusted within the ego BEV frame before fusion.}

{
Although the warping operation compensates the main feature shift caused by
staleness, the warped feature may contain local discrepancies caused by
bilinear sampling, imperfect displacement prediction, and spatially varying
object motion. Therefore, we further adopt a temporal adaptive
adjustment term to update the warped feature, given by
\begin{equation}
	\Delta\boldsymbol{F}_{l}^{\mathrm{adj}}(t_f)
	=
	\mathcal{C}_{\mathrm{adj}}
	\left(
	[
	\bar{\boldsymbol{F}}_{l}(t_f);
	\boldsymbol{a}_{lm}(t_f)
	]
	\right),
	\label{eq:feature-adjustment}
\end{equation}
where $\mathcal{C}_{\mathrm{adj}}(\cdot)$ is a compact convolutional block.}

{
To avoid uniformly applying this adjustment to all BEV regions, we compute a
spatial gate as
\begin{equation}
	\boldsymbol{g}_{l}
	=
	\sigma
	\left(
	\mathcal{C}_{\mathrm{gate}}
	\left(
	\Delta\boldsymbol{F}_{l}^{\mathrm{adj}}(t_f)
	\right)
	\right),
	\label{eq:adjustment-gate}
\end{equation}
where $\sigma(\cdot)$ is the sigmoid function, and
$\mathcal{C}_{\mathrm{gate}}(\cdot)$ is a lightweight convolutional layer that
outputs a spatial gate map. The final compensated feature is then obtained as
\begin{equation}
	\hat{\boldsymbol{F}}_{l}(t_f)
	=
	\mathcal{C}_{\mathrm{out}}
	\left(
	\bar{\boldsymbol{F}}_{l}(t_f)
	+
	\boldsymbol{g}_{l}
	\odot
	\Delta\boldsymbol{F}_{l}^{\mathrm{adj}}(t_f)
	\right),
	\label{eq:compensated-feature}
\end{equation}
where $\odot$ denotes element-wise multiplication, and
$\mathcal{C}_{\mathrm{out}}(\cdot)$ is a lightweight Conv2D layer. This design keeps the warped neighbor feature as the source representation, while allowing the compensation to adapt to local temporal shifts before RoI-level fusion.}\footnote{ {The feature staleness considered in this work is constructed from recorded frames in the public collaborative perception dataset, whose adjacent frames are separated by about $0.1$~s. Therefore, the proposed compensation is designed for short-term   staleness. Extremely abrupt	maneuvers or long-delay cases may require additional motion modeling, which is left for future work.}}

\subsubsection{Alignment Objective}
\label{sec:align-objective}
{ During training, the alignment module is supervised using the fusion-time feature of the same neighbor. For neighbor $l$, the projected asynchronous feature is used as the input, while the feature extracted from the same neighbor at $t_f$ is transformed to the ego fusion frame and used as the training target. This target is constructed only from the training data and is not available during inference. Since the target comes from the same neighbor rather than from the ego vehicle, the supervision encourages the compensated neighbor feature to approach its own fusion-time representation instead of imitating the ego observation. }

{ Let $\boldsymbol{F}_{l}^{\mathrm{cur}}(t_f)$ denote this target feature. The alignment loss is defined as 
\begin{equation} 
	\mathcal{L}_{\mathrm{align}} = \frac{1}{|\Omega_l^{\mathrm{obj}}|} \sum_{\boldsymbol{\upsilon}\in\Omega_l^{\mathrm{obj}}} \mathrm{SmoothL1} \left( \hat{\boldsymbol{F}}_l(t_f,\boldsymbol{\upsilon}) - \boldsymbol{F}_{l}^{\mathrm{cur}}(t_f,\boldsymbol{\upsilon}) \right),
	\label{eq:comp-loss} 
\end{equation} 
where $\Omega_l^{\mathrm{obj}}\subseteq\Omega_{\mathrm{BEV}}$ is the object-related BEV locations derived from ground-truth boxes during training. $\mathrm{SmoothL1}(\cdot)$ is the robust regression loss widely used in object detection~\cite{girshick2015fast}. Computing the loss on object-related locations reduces the dominance of background regions and focuses the supervision on motion-sensitive regions. When multiple neighbors are available, $\mathcal{L}_{\mathrm{align}}$ is averaged over the corresponding neighboring features. }

\subsection{Weighted Feature Fusion}
\label{sec:weighted-fusion}

After spatiotemporal feature alignment, vehicle $m$ obtains  a set of  compensated
neighbor features $\{\hat{\boldsymbol{F}}_l(t_f)\}_{l\in{L}_m}$ in the
ego fusion frame. Motivated by ETSI Collective Perception Service and recent
vehicular perception studies~\cite{etsi_ts103324,lyu2025accuracy,
wolff2025uncertainty}, we assign an RoI-level fusion utility to each candidate
RoI from each neighboring vehicle, so as to evaluate its usefulness at fusion
time $t_f$.

\subsubsection{RoI-level Fusion Utility} 
\label{sec:fusion-utility}
{For neighboring vehicle $l$ and RoI $r$, we compute an RoI-level fusion utility $\mathcal{V}_{lm}(r,t_f)$, which jointly considers  freshness, synchronization reliability, and content complementarity.}

Specifically, the freshness utility is defined based on the delivery-time AoI:
\begin{equation}
U_{\mathrm{fresh}}(r,t_f) = \exp \left( -\lambda \tilde{\mathcal{A}}_{lm}(r,t_f) \right), 
\label{eq:fresh-utility} 
\end{equation} 
where $\lambda>0$ is an AoI decay parameter. This utility decreases with delivery-time AoI, so older RoI features receive lower freshness utility.

The reliability utility is defined using the synchronization reliability factor
derived in Sec.~\ref{sec:temporal-uncertainty}:
\begin{equation}
U_{\mathrm{rel}}(r,t_f)
=
c_{lm}(r,t_f),
\label{eq:rel-utility}
\end{equation}
where RoIs with larger residual synchronization uncertainty receive smaller
reliability utility.

{Although freshness and reliability indicate whether the neighbor feature is timely and properly synchronized, they do not measure whether RoI $r$ provides object-related evidence  that complements the ego observation. We therefore use objectness to quantify content complementarity. Let $\boldsymbol{O}_l(\boldsymbol{\upsilon})$ and $\boldsymbol{O}_m(\boldsymbol{\upsilon})$ denote the objectness logits predicted from the compensated neighbor feature $\hat{\boldsymbol{F}}_l(t_f)$ and the ego feature $\boldsymbol{F}_m(t_f)$, respectively. The objectness is obtained by averaging the objectness probabilities within each RoI: 
\begin{align} 
	o_\ell(r) = \frac{1}{|\Omega_r|} \sum\nolimits_{\boldsymbol{\upsilon}\in\Omega_r} \sigma \left( \boldsymbol{O}_\ell(\boldsymbol{\upsilon}) \right), \  \ell\in\{m,l\}.
	\label{eq:roi-objectness} 
\end{align} }

{Given  the neighbor and ego objectness in RoI $r$, $o_l(r)$ and $o_m(r)$, the content complementarity is defined as 
\begin{equation} 
	\nu_{lm}(r) = o_l(r) \left( 1-o_m(r) \right), 
	\label{eq:objectness-complementarity} 
\end{equation}
emphasizing RoIs where the neighbor feature has high objectness while the ego feature has weak objectness. Therefore, $\nu_{lm}(r)$ favors object-related information that is complementary to, rather than redundant with, the ego observation. }

{
The content utility is then defined as
\begin{equation}
	U_{\mathrm{cont}}(r,t_f)
	=
	1
	+
	\beta_{\nu}
	\operatorname{clip}
	\left(
	\nu_{lm}(r),0,\nu_{\max}
	\right),
	\label{eq:cont-utility}
\end{equation}
where $\beta_{\nu}>0$ controls the contribution of content complementarity, and
$\nu_{\max}$ bounds this contribution. The constant $1$ keeps the content term
positive, while the clipping operation prevents large complementarity values
from dominating the overall utility.
}

{
The three utility terms are combined as
\begin{equation}
	U_{\mathrm{sum}}(r,t_f)
	=
	w_f U_{\mathrm{fresh}}(r,t_f)
	+
	w_r U_{\mathrm{rel}}(r,t_f)
	+
	w_c U_{\mathrm{cont}}(r,t_f),
	\label{eq:utility-comb}
\end{equation}
where $w_f,w_r,w_c\ge 0$ control the contributions of freshness, reliability,
and content complementarity, respectively. In implementation, these weights are
generated from trainable parameters through a softplus mapping and are
initialized equally. Therefore, the relative importance of the three utility
terms is learned rather than manually fixed.
}

{
The final RoI-level fusion utility is further modulated according to the joint values of the three utility terms:
\begin{align} 
	\mathcal{V}_{lm}(r,t_f) = U_{\mathrm{sum}}(r,t_f) \cdot \Psi \big(& U_{\mathrm{fresh}}(r,t_f), U_{\mathrm{rel}}(r,t_f), \notag\\ &U_{\mathrm{cont}}(r,t_f) \big), \label{eq:fusion-utility-score} 
\end{align} 
where $\Psi(\cdot)$ is a lightweight MLP that outputs a positive modulation factor. In this formulation, $U_{\mathrm{sum}}$ represents the learned weighted combination of the three utility terms, while $\Psi(\cdot)$ adjusts their joint influence on the final fusion utility. Therefore, the final RoI-level utility is not fixed by a hand-crafted weighting rule, while moderately stale but complementary RoIs can remain useful for fusion. }

\subsubsection{Utility-Based RoI Selection}
\label{sec:roi-select}

Based on the RoI-level fusion utility, the ego vehicle retains the RoIs whose utility is no smaller than a threshold:
\begin{equation}
\mathcal{R}_{lm}(t_f)
=
\left\{
r\in\mathcal{R}
\mid
\mathcal{V}_{lm}(r,t_f) \ge \xi
\right\},
\  l\in{L}_m,
\label{eq:roi_selection}
\end{equation}
where $\xi>0$ controls the selectivity of RoI fusion. A larger $\xi$ keeps only RoIs with higher utility, while a smaller $\xi$ allows more candidate RoIs to be retained.  

{ 
The retained RoIs are then organized as a feature-level CPM, which specifies the RoI features and metadata made available from vehicle $l$ to vehicle $m$ for fusion. Following the ETSI Collective Perception Service and recent studies on object prioritization in vehicular perception~\cite{etsi_ts103324,lyu2025accuracy,wolff2025uncertainty}, we write
\begin{equation}
	\mathcal{M}_{lm}^{\mathrm{CPM}}(t_f)
	=
	\left\{
	\left(
	s_{lm}(r),
	\left.\boldsymbol{F}_l\right|_{\Omega_r},
	o_l(r)
	\right)
	\,\middle|\,
	r\in\mathcal{R}_{lm}(t_f)
	\right\},
	\label{eq:fcpm}
\end{equation}
where $s_{lm}(r)$ is the synchronized feature-generation time of RoI $r$,
$\left.\boldsymbol{F}_l\right|_{\Omega_r}$ is the RoI feature over $\Omega_r$, and $o_l(r)$ is its neighbor-side objectness.
}

\subsubsection{RoI-Weighted Feature Fusion}
After RoI selection, only the retained neighbor RoIs are used for fusion. For a selected RoI $r$, let ${L}_m(r,t_f)$ be the set of neighboring vehicles
that retain this RoI, i.e., ${L}_m(r,t_f)
=\{l\in{L}_m \mid r\in\mathcal{R}_{lm}(t_f)\}$.
The RoI-level utilities are normalized over these neighbors to obtain the
RoI-level fusion weight:
\begin{equation}
\alpha_{lm}(r,t_f)
=
\frac{
	\exp\left(\mathcal{V}_{lm}(r,t_f)/T_{\alpha}\right)
}{
	\sum_{l'\in {L}_m(r,t_f)}
	\exp\left(\mathcal{V}_{l'm}(r,t_f)/T_{\alpha}\right)
	+
	\epsilon
},
\label{eq:utility-softmax}
\end{equation}
where $T_{\alpha}>0$ is a temperature parameter and $\epsilon$ is a small
constant for numerical stability. Thus, $\alpha_{lm}(r,t_f)$ represents the
relative contribution of neighbor $l$ to RoI $r$ among the retained neighboring
features.

Since fusion is performed on the BEV grid, the RoI-level weights are expanded
to a BEV-grid weight map $\boldsymbol{\alpha}_{lm}(t_f)$ by assigning
$[\boldsymbol{\alpha}_{lm}(t_f)]_{\boldsymbol{\upsilon}}
=\alpha_{lm}(r,t_f)$ for $\boldsymbol{\upsilon}\in\Omega_r$ and
$r\in\mathcal{R}_{lm}(t_f)$, and 0 otherwise.

Before aggregation, each compensated neighbor feature is projected by a lightweight $1\times1$ convolution:
\begin{equation}
\boldsymbol{F}_{l}^{\mathrm{proj}}(t_f)
=
\mathcal{W}_{\mathrm{fuse}}
\left(
\hat{\boldsymbol{F}}_l(t_f)
\right),
\label{eq:fusion-proj}
\end{equation}
where $\mathcal{W}_{\mathrm{fuse}}(\cdot)$   maps the compensated neighbor feature to the fusion feature
space. Finally, the fused feature is obtained
\begin{equation}
\boldsymbol{F}_{m}^{\mathrm{fuse}}(t_f)
=
\Gamma
\left(
\boldsymbol{F}_m(t_f)
+
\sum_{l\in{L}_m}
\boldsymbol{\alpha}_{lm}(t_f)
\odot
\boldsymbol{F}_{l}^{\mathrm{proj}}(t_f)
\right),
\label{eq:utility-attention-fusion}
\end{equation}
where $\Gamma(\cdot)$ is a convolutional aggregation module, and
$\boldsymbol{\alpha}_{lm}(t_f)$ is broadcast along the channel dimension when
multiplied with $\boldsymbol{F}_{l}^{\mathrm{proj}}(t_f)$.

\subsubsection{Training Objective}

The fused BEV feature is supervised by an anchor-based detection objective
following PointPillars-based detectors~\cite{lang2019pointpillars}:
\begin{equation}
\mathcal{L}_{\mathrm{det}}
=
\mathcal{L}_{\mathrm{cls}}
+
\lambda_{\mathrm{reg}}\mathcal{L}_{\mathrm{reg}}
+
\lambda_{\mathrm{dir}}\mathcal{L}_{\mathrm{dir}},
\label{eq:det-loss}
\end{equation}
where $\mathcal{L}_{\mathrm{cls}}$ is the sigmoid focal classification
loss~\cite{lin2017focal}, $\mathcal{L}_{\mathrm{reg}}$ is the SmoothL1
bounding-box regression loss, and $\mathcal{L}_{\mathrm{dir}}$ is the direction
classification loss. We set $\lambda_{\mathrm{reg}}=2.0$ and
$\lambda_{\mathrm{dir}}=0.2$.

Since the content utility depends on objectness, we also supervise an objectness head using binary cross entropy, $\mathcal{L}_{\mathrm{obj}}$, where its objectness target  is obtained by projecting ground-truth boxes onto the BEV grid.

The overall training objective is
\begin{equation}
\mathcal{L}
=
\mathcal{L}_{\mathrm{det}}
+
\lambda_{\mathrm{obj}}\mathcal{L}_{\mathrm{obj}}
+
\eta_{\mathrm{align}}\mathcal{L}_{\mathrm{align}},
\label{eq:overall-loss}
\end{equation}
where the objectness loss supports content evaluation, while the alignment loss provides supervision for feature compensation. In our simulations, we set $\lambda_{\mathrm{obj}}=0.5$ and
$\eta_{\mathrm{align}}=0.2$. 

\section{Evaluations}
In this section, we compare the proposed framework against representative baselines for collaborative perception.
\subsection{Simulation Setup}
\label{sec:sim-setup}

\subsubsection{Dataset} 
{We evaluate the proposed framework on LiDAR-based 3D object detection using the V2X-Sim dataset~\cite{li2022v2xsim,v2xsim_project}. }  V2X-Sim is generated with SUMO~\cite{krajzewicz2012recent} and CARLA~\cite{dosovitskiy2017carla}, and provides multi-agent LiDAR point clouds and 3D bounding-box annotations. It contains 100 scenes, with 70 scenes for training, 15 scenes for validation, and 15 scenes for testing. Each scene contains up to six agents, including vehicles and one RSU. Agent~1 is treated as the ego agent, and the remaining agents are treated as neighboring agents.

The raw point clouds are cropped to the perception range $[-32,-32,-3,32,32,2]$~m. For the PointPillars backbone, the point clouds are voxelized into pillars with voxel size $(0.25,0.25,5.0)$~m and encoded into BEV features. The ego and neighboring BEV features are then used for spatiotemporal faeture alignment,  weighted fusion, and final detection.

\subsubsection{Data Generation and Metadata Construction}
\label{sec:data-gen}
For each scene and each agent, we generate the synchronization and communication metadata required by the proposed framework.

\paragraph{Network Synchronization}

To emulate unsynchronized clocks, each neighboring agent $l$ is assigned a raw clock model $ \tau_l^{\mathrm{raw}}(t) = t+\theta_l+\varpi_l t+\varepsilon_l(t), $ where the clock offset follows $\theta_l\sim\mathcal{U}(-500,500)$~ms and the clock skew follows $\varpi_l\sim\mathcal{N}(0,10~\mathrm{ppm})$. For the ego agent $m$ and neighboring agent $l$, we simulate 10 rounds of two-way timestamp exchange and apply R3SRKF to estimate the relative clock state $(\hat{\theta}_{lm,k|k},\hat{\varpi}_{lm,k|k})$ and the synchronization uncertainty.

The base delay is set to $\Delta t=2$ frames, and the frame interval is 0.1~s. For each neighboring input, the actual stale frame is determined by the base delay together with the frame offsets induced by residual clock error and communication delay. The generation timestamp of such neighbor frame is mapped to the ego temporal reference, and the delivery-time AoI is computed relative to the ego fusion frame.

\paragraph{Communication Conditions}

{The communication settings instantiate the NR-V2X sidelink model described in Sec.~\ref{subsec:comm_model}. The configuration follows the NR sidelink numerology and physical-channel specifications in~\cite{3gpp_ts38211,3gpp_ts38214},	while Mode~2 autonomous resource access is configured according to~\cite{3gpp_tr37885,ali2021nr_v2x_mode2}. The main parameters are summarized in Table~\ref{tab:nr_v2x_params}.}

{We use the ns-3 5G-LENA NR V2X module~\cite{patriciello2019e2e,ali2021nr_v2x_mode2} to generate packet reception records under sensing-based Mode~2 sidelink operation. Based on these records, we fit the PSSCH BLER with a logistic function of the effective SINR:
\begin{equation}
	\Lambda(\gamma_{\mathrm{eff}})
	=
	\frac{1}{1+\exp\left(\alpha_{\Lambda}
		(\gamma_{\mathrm{eff}}-\gamma_0)\right)},
	\label{eq:bler_ns3}
\end{equation}
where $\Lambda(\gamma_{\mathrm{eff}})$ denotes the PSSCH BLER,
$\gamma_0=6.57$~dB, and $\alpha_{\Lambda}=1.155$. Accordingly,
$1-\Lambda(\gamma_{\mathrm{eff}})$ gives the packet decoding probability used in the link-abstraction model.
}

{
Based on this link abstraction, we consider favorable, moderate, and congested communication conditions to examine the proposed collaborative perception framework under different levels of channel quality and Mode~2 access latency, and the moderate condition is used as the default setting. For each neighbor-to-ego sidelink, the SINR is sampled as
$\gamma_{\mathrm{SINR,dB}}\sim\mathcal{N}(\mu_\gamma,\sigma_\gamma^2)$, where the condition-specific values of $(\mu_\gamma,\sigma_\gamma)$ are listed in Table~\ref{tab:comm_conditions}. The sampled SINR determines the PSSCH BLER through~\eqref{eq:bler_ns3}. The Mode~2 access latency is generated according to~\eqref{eq:access_latency}, using the access-latency window $[T_1,T_2]$ of the corresponding condition. For RoI-payload transmission, the payload capacity is computed from the configured subchannel resource size and the spectral efficiency of the selected MCS, and the corresponding transmission delay is obtained from \eqref{eq:rate_based_delivery_time}. Finally, the total communication delay used for delivery-time AoI is computed by \eqref{eq:comm_delay_total}. 
}

\begin{table}[t] 
\centering \caption{NR-V2X sidelink  simulation parameters.} \label{tab:nr_v2x_params} 
\begin{tabularx}{\columnwidth}{ 
		>{\raggedright\arraybackslash}p{0.44\columnwidth} 
		>{\raggedright\arraybackslash}X} \hline 
	\textbf{Parameter} & \textbf{Value} \\ \hline
	Carrier frequency & 5.89 GHz \\ Channel bandwidth & 20 MHz \\
	Subcarrier spacing & 30 kHz \\ 
	Slot duration & 0.5 ms \\ 
	Number of subchannels & 2 \\ 
	RoI payload size & 3 KB\\ 
	MCS candidates & QPSK 1/2, QPSK 3/4, 16QAM 1/2, 16QAM 3/4 \\ \hline 
\end{tabularx} 
\end{table}

\begin{table}[t]  
\centering 
\caption{Communication condition settings.} 
\label{tab:comm_conditions} 
\begin{tabular}{l c c } \hline 
	\textbf{Condition} & \textbf{SINR (dB)} & \textbf{$[T_1,T_2]$ (ms)}  \\ \hline 
	Favorable & $\mathcal{N}(14,5^2)$ & $[5,60]$  \\ 
	Moderate & $\mathcal{N}(9,6^2)$ & $[10,120]$  \\ 
	Congested & $\mathcal{N}(4,7^2)$ & $[20,250]$  \\ \hline 
\end{tabular} \end{table}

\subsubsection{Evaluation Metric}
\label{sec:eval-metric}
{Following the collaborative perception benchmarks~\cite{9307334,wei2023asynchrony}, we evaluate perception performance using mean average precision (mAP). Specifically, mAP@0.5 and mAP@0.7 denote mAP evaluated at BEV intersection-over-union (IoU) thresholds $\tau_{\mathrm{IoU}}=0.5$ and $\tau_{\mathrm{IoU}}=0.7$, respectively. The IoU is computed on the BEV plane between the predicted and ground-truth bounding boxes. A prediction is counted as a true positive (TP) if its IoU with an unmatched ground-truth box in the same frame is no smaller than $\tau_{\mathrm{IoU}}$; otherwise, it is counted as a false positive (FP).
Precision and recall are computed as
\begin{equation} 
	\mathrm{Prec} = \frac{\mathrm{TP}} {\mathrm{TP}+\mathrm{FP}}, \quad \mathrm{Rec} = \frac{\mathrm{TP}} {N_{\mathrm{gt}}}, 
	\label{eq:precision-recall} 
\end{equation} 
where $N_{\mathrm{gt}}$ is the total number of ground-truth objects. The average precision (AP) is computed as the area under the precision--recall curve. The mAP is obtained by averaging AP over the evaluated object categories: 
\begin{equation} 
	\mathrm{mAP}(\tau_{\mathrm{IoU}}) = \frac{1}{|\mathcal{C}_\text{cat}|} \sum_{c\in\mathcal{C}_\text{cat}} \mathrm{AP}_{c}(\tau_{\mathrm{IoU}}), 
	\label{eq:map} 
\end{equation} 
where $\mathcal{C}_\text{cat}$ is the set of object categories. }

All models are trained using the Adam optimizer with a batch size of~2 and an initial learning rate of $2\times10^{-3}$. The learning rate is scheduled by cosine annealing over 20 epoches.  The simulations are implemented in PyTorch and conducted on NVIDIA RTX A6000 GPUs.
\subsection{Comparisons of Collaborative Perception Algorithms}
We compare the performance of our proposed spatiotemporal feature alignment and weighted fusion algorithm with baselines, as described below.
\begin{itemize}
\item \textbf{SyncNet~\cite{lei2022latency}} uses a latency compensation module with feature-attention symbiotic estimation and time modulation to adapt asynchronous perceptual features to a common timestamp before fusion.
\item \textbf{CoBEVFlow~\cite{wei2023asynchrony}} uses BEV flow to compensate temporal asynchrony by reassigning asynchronous perceptual features to appropriate BEV locations, and can handle irregular continuous timestamps.
\end{itemize}

\begin{figure}[!t]
\centering
\subfloat[mAP@0.5.]{
	\includegraphics[width=0.47\linewidth]{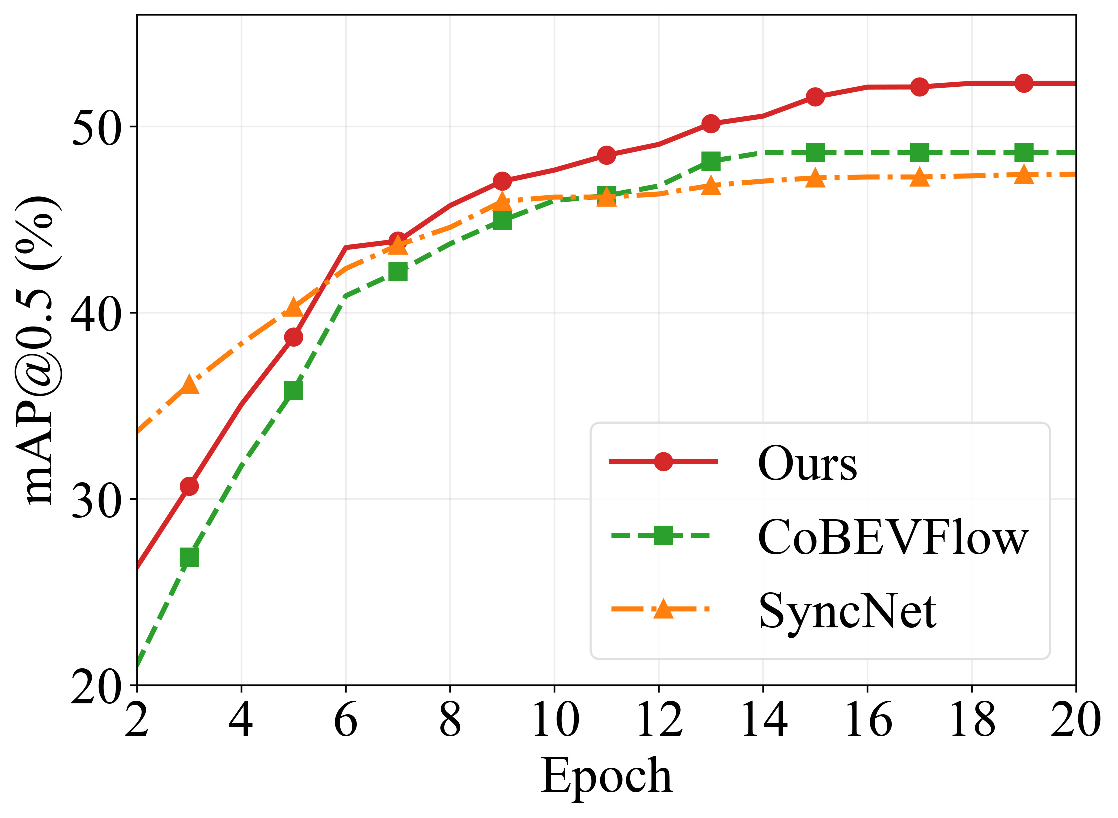}
}
\subfloat[mAP@0.7.]{
	\includegraphics[width=0.47\linewidth]{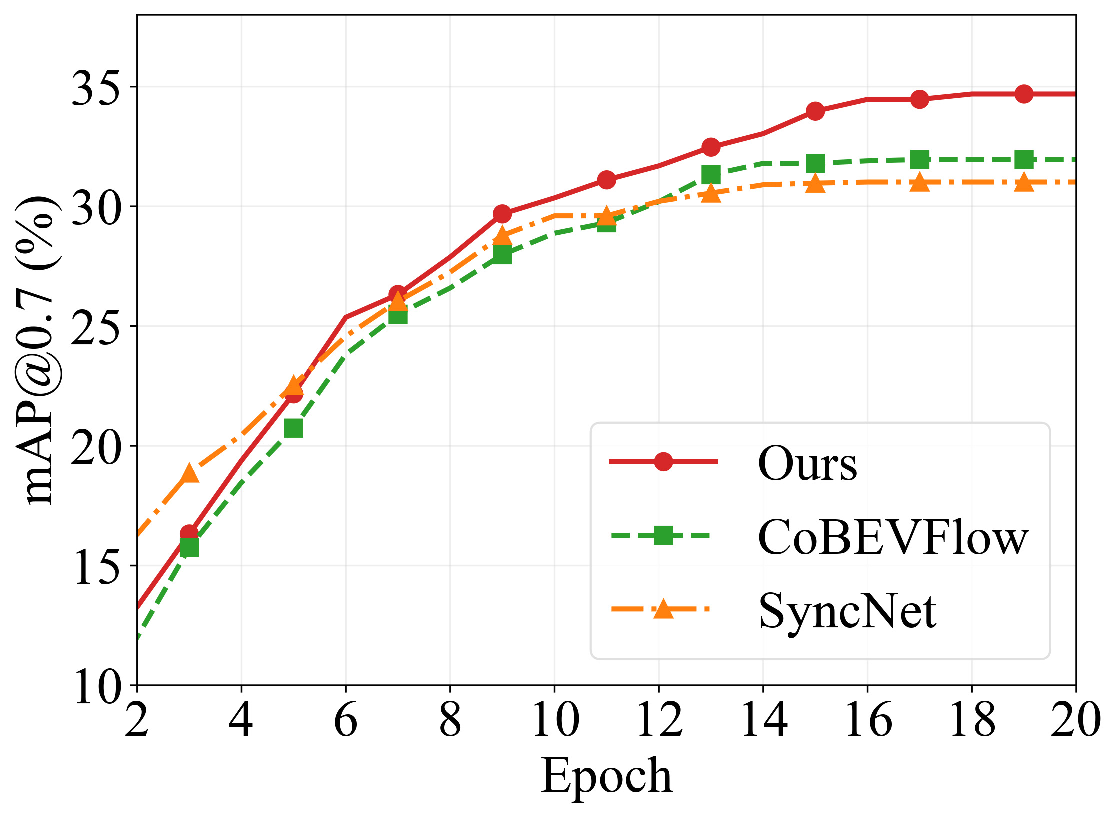}
}
\caption{Performance comparison of various schemes, including  our proposed method, CoBEVFlow, and SyncNet, evaluating their (a) mAP@0.5, and (b)  mAP@0.7.}
\label{fig:method}
\end{figure}

Fig.~\ref{fig:method} shows the perception performance of the three methods 
under moderate communication condition. For mAP@0.5, SyncNet achieves 
higher accuracy in the early epochs, while our method starts from a lower value 
but improves more rapidly.  After several 
training epochs, our method surpasses both baselines and maintains a consistent advantage in the later stage,  eventually reaching about $54\%$, compared with about $50\%$ for CoBEVFlow  and $49\%$ for SyncNet.
The advantage is more evident under the stricter mAP@0.7 criterion. Our method 
continues to improve after the early epochs and finally reaches about $36\%$, 
whereas CoBEVFlow and SyncNet converge to about $34\%$ and $32\%$, respectively.

This improvement is attributed to the delivery-time AoI, which evaluates the 
freshness of shared RoIs at their expected fusion time rather than relying only 
on the source timestamp or a fixed-delay assumption. Meanwhile, the RoI-level 
weighted fusion suppresses stale, temporally unreliable, or redundant neighboring 
features, preventing low-value shared information from degrading the fused 
representation.

\begingroup

\subsection{Impact of Communication Conditions}

To evaluate robustness under different NR-V2X sidelink conditions, we compare the perception performance under favorable, moderate, and congested settings.

As shown in Table~\ref{tab:comm_sensitivity}, the proposed method degrades only slightly as the communication condition worsens. Its mAP@0.5 decreases from $54.4\%$ to $53.6\%$ and $52.6\%$, while mAP@0.7 decreases from $37.3\%$ to $36.5\%$ and $35.3\%$, respectively. This shows that the proposed framework is robust to variations in channel quality, resource-selection latency, and packet delivery reliability.

Compared with CoBEVFlow and SyncNet, the proposed method consistently achieves higher accuracy under all communication conditions. Under the congested setting, it still obtains $52.6\%$ mAP@0.5 and $35.3\%$ mAP@0.7, outperforming CoBEVFlow by $3.9\%$ and $2.3\%$, and SyncNet by $5.6\%$ and $5.2\%$, respectively. This advantage comes from delivery-time freshness and RoI-level fusion utility, which 
down-weight stale or unreliable features before fusion and thereby improve robustness under time-varying V2X communication conditions.

\begin{table}[t]
\centering

\caption{Perception performance under different communication conditions.}
\label{tab:comm_sensitivity}
\setlength{\tabcolsep}{5pt}
\begin{tabular}{l l c c c}
	\hline
	\textbf{Metric} & \textbf{Method}
	& \textbf{Favorable} & \textbf{Moderate} & \textbf{Congested} \\
	\hline
	\multirow{3}{*}{mAP@0.5 (\%)}
	& Ours       & \textbf{54.4} & \textbf{53.6} & \textbf{52.6} \\
	& CoBEVFlow  & 50.7          & 49.7          & 48.7          \\
	& SyncNet    & 50.3          & 48.6          & 47.0          \\
	\hline
	\multirow{3}{*}{mAP@0.7 (\%)}
	& Ours       & \textbf{37.3} & \textbf{36.5} & \textbf{35.3} \\
	& CoBEVFlow  & 34.4          & 33.8          & 33.0          \\
	& SyncNet    & 32.9          & 31.5          & 30.1          \\
	\hline
\end{tabular}
\end{table}
\endgroup

\begin{table}[!t]
\centering

\caption{Sensitivity of voxel size and backbone.}
\label{tab:voxel_backbone_sensitivity}
\setlength{\tabcolsep}{5pt}
\begin{tabular}{l c c}
	\hline
	\textbf{Backbone and voxel size}  
	& \textbf{mAP@0.5 (\%)} & \textbf{mAP@0.7 (\%)} \\
	\hline
	PointPillars, 0.16 m & 39.0 & 16.3 \\
	\textbf{PointPillars, 0.25 m} & \textbf{53.6} & \textbf{36.5} \\
	PointPillars, 0.32 m & 45.4 & 30.2 \\
	SECOND, $(0.16,0.16,0.20)$ m & 47.8 & 30.0 \\
	\hline
\end{tabular}
\end{table}

% {
\subsection{Sensitivity of Voxel Size and Backbone}

{
We conduct two sensitivity studies under the moderate communication condition.
The first varies the xy-plane voxel size of the PointPillars encoder, while the
second replaces PointPillars with SECOND~\cite{yan2018second} to test whether
the proposed spatiotemporal alignment and RoI-level fusion design generalizes to
another feature extractor. PointPillars produces 2D BEV features from vertical
pillars, whereas SECOND extracts 3D voxel features before forming the feature
map used for alignment and fusion.
}

{
Table~\ref{tab:voxel_backbone_sensitivity} illustrates that the PointPillars with a
$0.25$~m voxel size achieves the best overall performance. A larger voxel size
($0.32$~m) produces a coarser BEV representation and loses spatial feature details,
whereas a smaller voxel size ($0.16$~m) increases the feature-map size and
computational burden. These results show a trade-off among feature resolution,
computational cost, and the effectiveness of the proposed alignment and fusion
framework.
}

{
With the SECOND backbone, the proposed framework achieves $47.8\%$ mAP@0.5 and
$30.0\%$ mAP@0.7. Although this result is lower than the best PointPillars
configuration, it shows that the proposed spatiotemporal alignment and
RoI-level fusion modules can also operate with voxel-based feature representations.
This supports the applicability of the proposed design beyond a single
PointPillars configuration.
}

\begin{figure}[!t]
\centering
\subfloat[mAP@0.5.]{
	\includegraphics[width=0.47\linewidth]{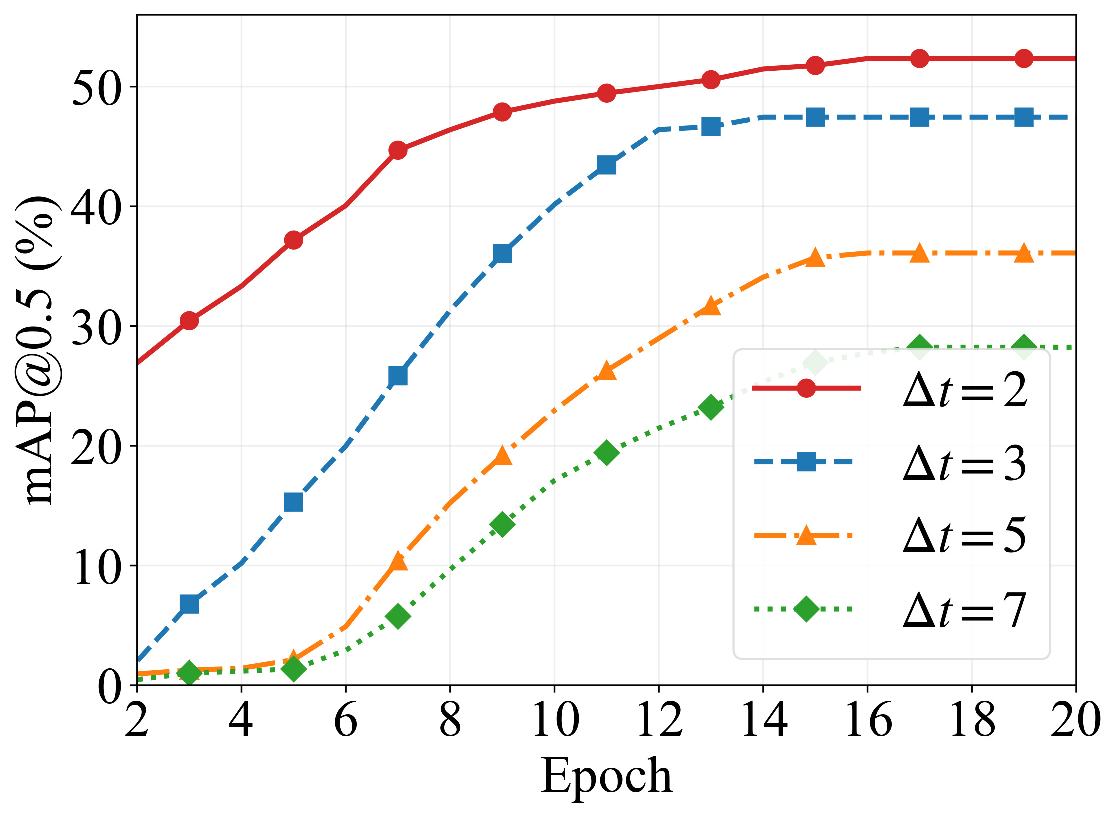}
}
\subfloat[mAP@0.7.]{
	\includegraphics[width=0.47\linewidth]{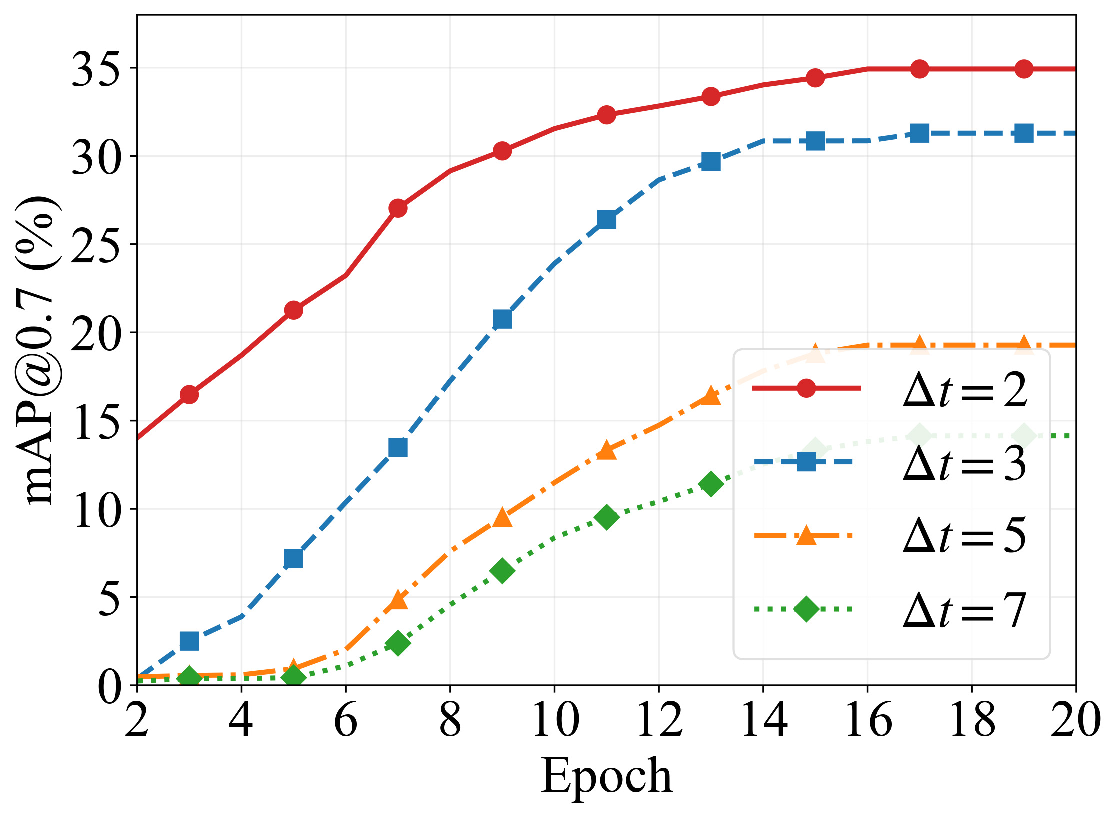}
}
\caption{Impact of temporal misalignment, evaluating their (a) mAP@0.5, and (b)  mAP@0.7.}
\label{fig:map_epoch}
\end{figure}
\begin{figure}[t]
\centering
\subfloat[$\Delta t$ = 2.]{
	\includegraphics[width=0.46\linewidth]{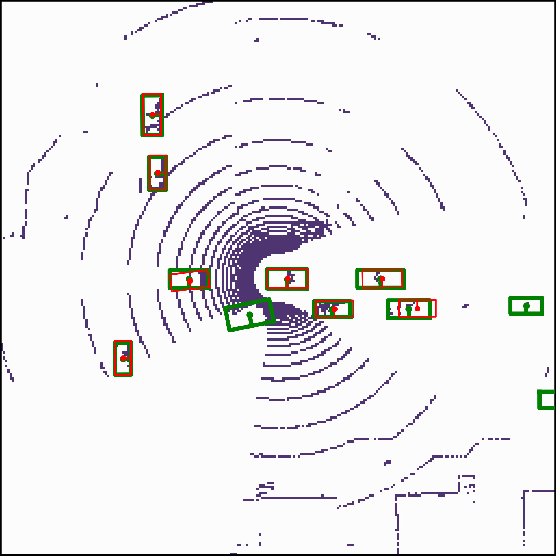}}
\hspace{5pt}
\subfloat[$\Delta t$ = 3.]{
	\includegraphics[width=0.46\linewidth]{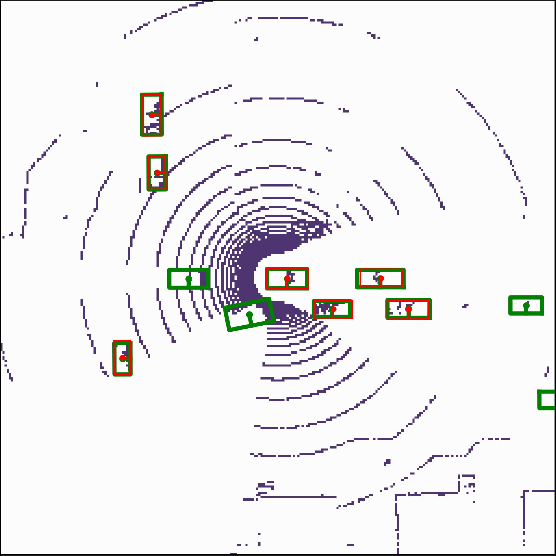}}\\
\subfloat[$\Delta t$ = 5.]{
	\includegraphics[width=0.46\linewidth]{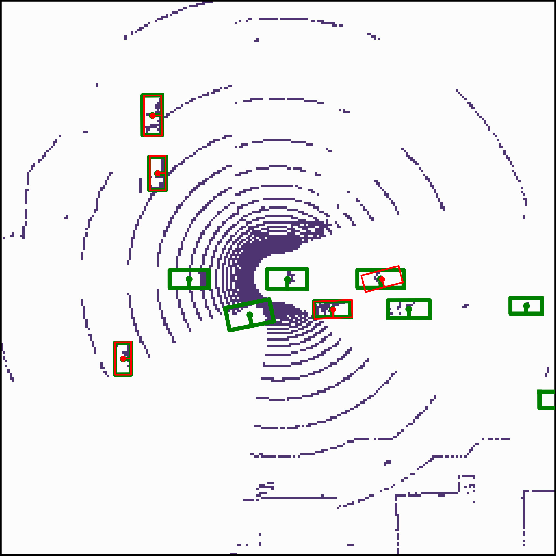}}
\hspace{5pt}
\subfloat[$\Delta t$ = 7.]{
	\includegraphics[width=0.46\linewidth]{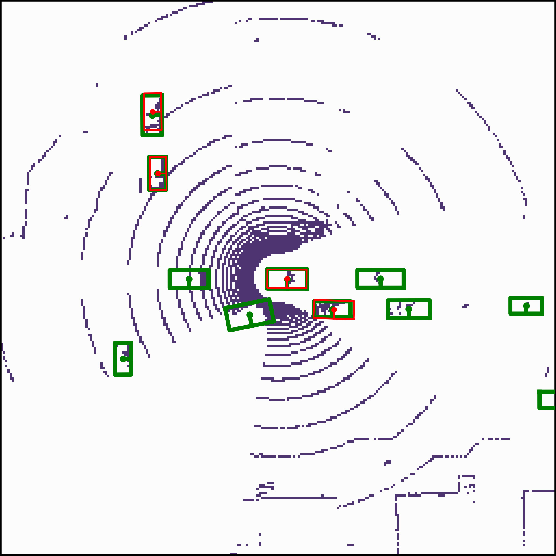}}
\caption{Visualization of temporal misalignment impact. \textcolor{green}{Green} boxes represent the ground truth, and \textcolor{red}{red} boxes are the  detected objects. 
}
\label{visualization}
\end{figure}

\subsection{Impact of Temporal Misalignment}
We further evaluate the sensitivity of the proposed framework to temporal
misalignment. The base delay is set to
$\Delta t\in\{2,3,5,7\}$ frames, while the delivery-time AoI also
includes the effects of residual clock error and communication delay. This
experiment examines how the proposed method performs as neighboring features
become increasingly stale and less consistent with the ego feature.

As shown in Fig.~\ref{fig:map_epoch}, all settings improve during training, but they converge to different performance levels. The cases $\Delta t=2,3$ converge faster and reach higher final accuracy, indicating that the proposed method can handle moderate temporal misalignment. When $\Delta t=5$, the model still converges, but the curve saturates at a lower plateau, around $46\%$ mAP@0.5 and $30\%$ mAP@0.7. When $\Delta t=7$, the final performance is much lower, especially under mAP@0.7, which indicates that large temporal gaps strongly affect precise perception.  {
This is mainly caused by the reduced effectiveness of temporal compensation	under large feature staleness. Since V2X-Sim is sampled at 10~Hz, $\Delta t\geq 5$ corresponds to at least $0.5$~s of base delay before considering residual clock offset and communication delay. At this time scale,  the stale neighbor feature differs from the current ego scene not only in spatial location, but also in occlusion state and object presence. Although our alignment design reduces predictable displacement errors, it cannot recover information that is absent or already outdated in the original neighbor feature.}

Fig.~\ref{visualization} also shows more shifted boxes and missed detections as	$\Delta t$ increases, which is consistent with the convergence curves in	Fig.~\ref{fig:map_epoch}. Overall, the proposed method is robust to moderate temporal misalignment, especially for $\Delta t\leq 3$, while larger delays	reduce the benefit of temporal compensation because the received features become	increasingly stale and inconsistent with the ego fusion time.

\subsection{Impact of Number of Agents}
{
To evaluate the scalability of the proposed framework with respect to vehicle density, we vary the total number of agents under different communication conditions. In addition to perception accuracy, we report the mean inference time to quantify the overhead introduced by the increasing agents.
}

\begin{table}[t]
\centering

\caption{Perception accuracy and inference time under different numbers of agents and communication conditions.}
\label{tab:agent_scalability}
\setlength{\tabcolsep}{5pt}
\begin{tabular}{l l c c c}
	\hline
	\textbf{Condition} & \textbf{Metric}
	& \textbf{2 Agents} & \textbf{4 Agents} & \textbf{6 Agents} \\
	\hline
	\multirow{3}{*}{Favorable}
	& mAP@0.5 (\%) & 37.5 & 47.5 & \textbf{54.4} \\
	& mAP@0.7 (\%) & 23.4 & 31.6 & \textbf{37.3} \\
	& Inference time (ms) & \textbf{21.2} & 51.5 & 80.5 \\
	\hline
	\multirow{3}{*}{Moderate}
	& mAP@0.5 (\%) & 37.6 & 46.8 & \textbf{53.6} \\
	& mAP@0.7 (\%) & 23.5 & 30.3 & \textbf{36.5} \\
	& Inference time (ms) & \textbf{21.1} & 51.5 & 80.5 \\
	\hline
	\multirow{3}{*}{Congested}
	& mAP@0.5 (\%) & 37.6 & 45.2 & \textbf{52.6} \\
	& mAP@0.7 (\%) & 23.6 & 29.0 & \textbf{35.3} \\
	& Inference time (ms) & \textbf{21.3} & 51.4 & 80.3 \\
	\hline
\end{tabular}
\end{table}

{
As shown in Table~\ref{tab:agent_scalability}, increasing the number of agents
consistently improves perception accuracy. Under the moderate communication
condition, mAP@0.5 increases from $37.6\%$ with 2 agents to $46.8\%$ with 4
agents and $53.6\%$ with 6 agents. Similarly, mAP@0.7 improves from $23.5\%$ to
$30.3\%$ and $36.5\%$, respectively. This confirms that additional agents
provide complementary observations and enlarge the effective perception
coverage, improving the fusion performance.
}

{This perception gain is accompanied by higher overhead. The inference time increases from about $21$~ms with 2 agents to about $51$~ms with 4 agents and $80$~ms with 6 agents. For a fixed number of agents, the inference time remains almost unchanged across favorable, moderate, and congested conditions, indicating that the end-to-end overhead is mainly determined by the number of agents. }

\subsection{Ablation Study}
To quantify the contribution of each component in the proposed framework, we conduct ablation studies by removing the freshness, reliability, and content-complementarity terms, and compare them with our full model.
\begin{itemize}
\item \textbf{w/o $U_{\mathrm{fresh}}$} removes the delivery-time freshness term, so the fusion weight does not account for the expected 
staleness of the RoI at the fusion time.
\item \textbf{w/o $U_{\mathrm{rel}}$} removes the reliability term associated with synchronization uncertainty, while retaining freshness and content complementarity.
\item \textbf{w/o $U_{\mathrm{cont}}$} removes the content-complementarity 
term and assigns fusion weights without evaluating whether a shared RoI 
provides additional perception evidence.
\end{itemize}
 
\begin{figure}[!t]
	\centering
	\subfloat[mAP@0.5.]{
		\includegraphics[width=0.47\linewidth]{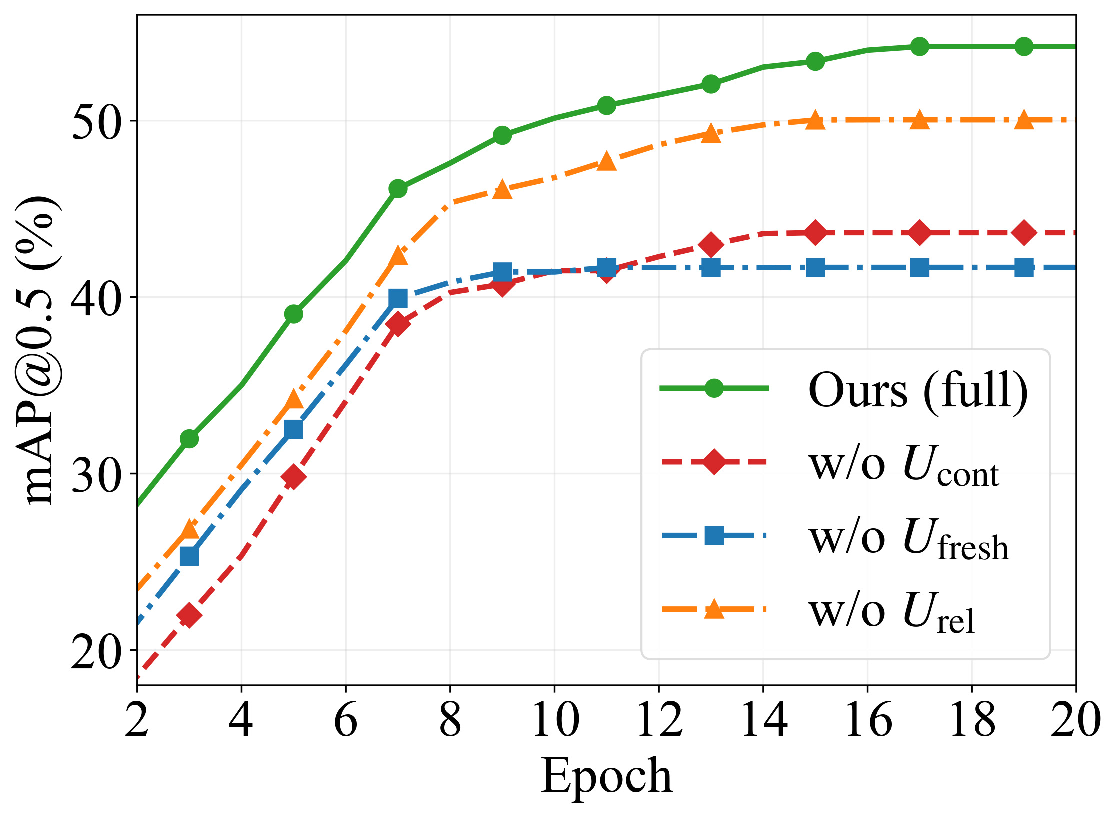}
		\label{fig:ablation_ap05}
	}
	\hfill
	\subfloat[mAP@0.7.]{
		\includegraphics[width=0.47\linewidth]{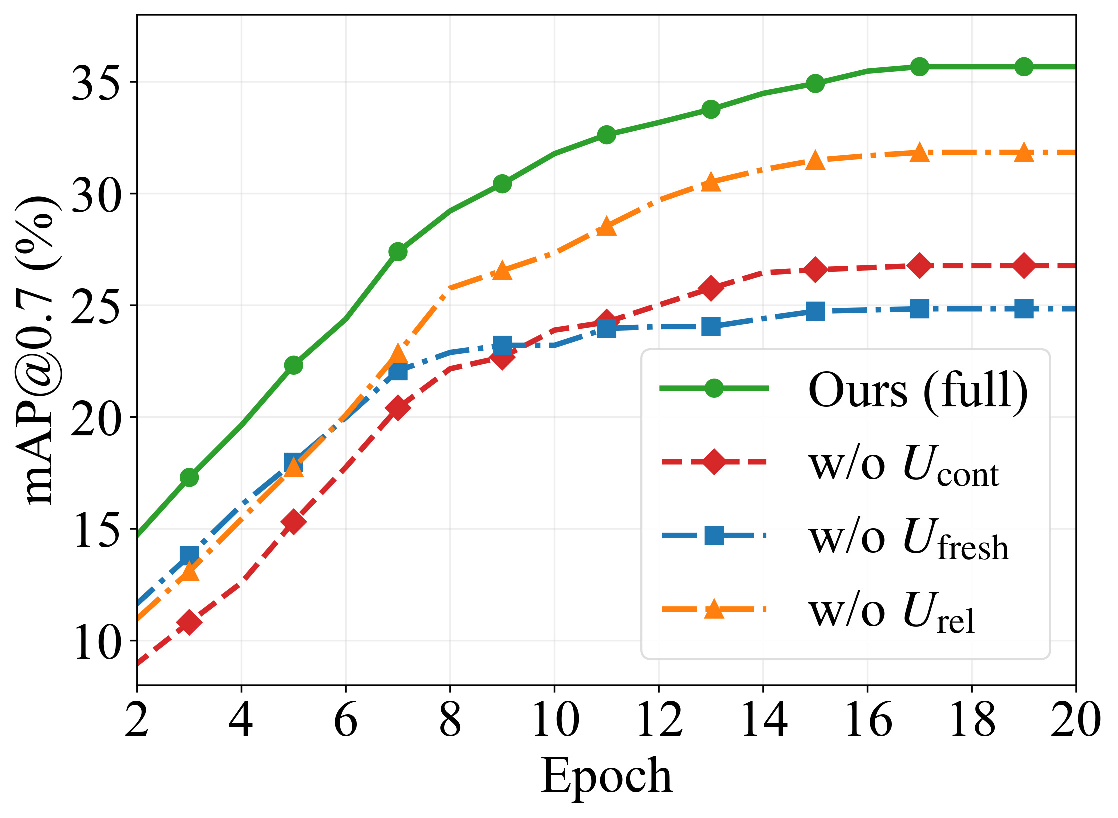}
		\label{fig:ablation_ap07}
	}
	\caption{Performance comparison of the ablation schemes in terms of
		(a) mAP@0.5 and (b) mAP@0.7.}
	\label{fig:ablation}
\end{figure}
Fig.~\ref{fig:ablation} indicates that the full model consistently achieves the best performance under both IoU thresholds. Removing $U_{\mathrm{fresh}}$ causes a clear performance drop, especially in the later epochs, indicating that delivery-time AoI is important for suppressing stale features before fusion. Removing $U_{\mathrm{cont}}$ also degrades the final accuracy, which 
shows that content complementarity is needed to avoid assigning high weights to redundant or weakly informative regions.
The scheme without $U_{\mathrm{rel}}$ performs better than the other two ablated schemes but remains below the full model. Together, these results indicate that the proposed framework benefits from
jointly considering multiple aspects of RoI usefulness, rather than relying on freshness, reliability, or content information alone.

\section{Conclusion}
This paper has introduced a spatiotemporal feature alignment and weighted fusion framework for collaborative perception enabled by network synchronization and AoI. To build a consistent temporal reference, this framework estimates clock states and maps  vehicle timestamps onto the ego temporal reference. Based on this, delivery-time AoI has been defined by jointly considering feature generation time and communication delay, so that the freshness of a candidate feature is derived at its expected fusion time.

For spatiotemporal alignment, geometric projection has been used to map neighbor features into the ego coordinate frame, while temporal feature compensation has been designed to update delayed features toward the fusion time. To improve perception efficiency and robustness, we have further
constructed an RoI-level weighted fusion scheme from delivery-time freshness, synchronization reliability, and content complementarity, reducing the influence of stale, uncertain, or weakly complementary regions during feature fusion.

Experiments under clock drift and communication delay have demonstrated that the proposed method consistently improves perception performance over representative baselines. The results have also shown its robustness under different communication conditions, temporal misalignment levels, and numbers
of agents, as well as its generality across backbone architectures and voxel sizes. Ablation studies have further confirmed the contributions of delivery-time AoI, synchronization reliability, and
content complementarity.

{
In future work, we will evaluate the proposed framework through full
protocol-stack network simulation and real V2X testbed experiments, examining the performance of the proposed framework under practical network dynamics, hardware clock behavior, and real timing variations.} Moreover, we will extend the framework toward multimodal collaborative perception with cameras, radar, etc., enabling shared information from heterogeneous sensing modalities to be fused according to AoI,  complementarity, or other impact factors.

\bibliographystyle{IEEEtran}
\bibliography{IEEE.bib}

\end{document}